\documentclass[twocolumn]{aastex631}
\usepackage{amsmath}
\usepackage[T1]{fontenc}
\usepackage{upgreek}
\usepackage{breqn}

\newcommand{\PSUAA}{Department of Astronomy \& Astrophysics, 525 Davey Laboratory, The Pennsylvania State University, University Park, PA, 16802, USA}
\newcommand{\PSUCEHW}{Center for Exoplanets and Habitable Worlds, 525 Davey Laboratory, The Pennsylvania State University, University Park, PA, 16802, USA}
\newcommand{\PSETI}{Penn State Extraterrestrial Intelligence Center, 525 Davey Laboratory, The Pennsylvania State University, University Park, PA, 16802, USA}
\newcommand{\UA}{Steward Observatory, The University of Arizona, 933 N.\ Cherry Ave, Tucson, AZ 85721, USA}

\newcommand{\Penn}{Department of Physics and Astronomy, University of Pennsylvania, 209 S 33rd St, Philadelphia, PA 19104, USA}

\newcommand{\STScI}{Space Telescope Science Institute, 3700 San Martin Dr, Baltimore, MD 21218, USA}
\newcommand{\JHU}{Department of Physics and Astronomy, Johns Hopkins University, 3400 N Charles St, Baltimore, MD 21218, USA}
\newcommand{\GoddardESAL}{Exoplanets and Stellar Astrophysics Laboratory, NASA Goddard Space Flight Center, Greenbelt, MD 20771, USA}

\newcommand{\NOAO}{NSF's National Optical-Infrared Astronomy Research Laboratory, 950 N.\ Cherry Ave., Tucson, AZ 85719, USA}

\newcommand{\Macquarie}{Department of Physics and Astronomy, Macquarie University, Balaclava Road, North Ryde, NSW 2109, Australia}

\newcommand{\JPL}{Jet Propulsion Laboratory, California Institute of Technology, 4800 Oak Grove Drive, Pasadena, California 91109}

\newcommand{\UCI}{Department of Physics \& Astronomy, The University of California, Irvine, Irvine, CA 92697, USA}
\newcommand{\Carleton}{Carleton College, One North College St., Northfield, MN 55057, USA}
\newcommand{\Carnegie}{Earth and Planets Laboratory, Carnegie Institution for Science, 5241 Broad Branch Road, NW, Washington, DC 20015, USA}
\newcommand{\PSUICS}{Institute for Computational and Data Sciences, The Pennsylvania State University, University Park, PA, 16802, USA}
\newcommand{\PSUCASt}{Center for Astrostatistics, 525 Davey Laboratory, The Pennsylvania State University, University Park, PA, 16802, USA}

\newcommand{\Princeton}{Department of Astrophysical Sciences, Princeton University, 4 Ivy Lane, Princeton, NJ 08540, USA}
\newcommand{\RUSSELL}{Henry Norris Russell Fellow}

\newcommand{\TIFR}{Department of Astronomy and Astrophysics, Tata Institute of Fundamental Research, Homi Bhabha Road, Colaba, Mumbai 400005, India}

\newcommand{\cms}{cm~s$^{-1}$}
\newcommand{\teff}{T_{\rm eff}}

\newcommand{\startable}{\ref{tab:stellar_params}}
\newcommand{\asterotable}{\ref{tab:astero_params}}
\newcommand{\sinc}{{\rm sinc}}

\newcommand{\rrev}{}
\newcommand{\rstrike}{\vphantom}

\newcommand{\rrevb}{}
\received{August 5, 2022}
\revised{September 29, 2022}
\accepted{September 30, 2022}
\submitjournal{AJ}

\shorttitle{HD 35833 p-mode Oscillations}
\shortauthors{Gupta et al.}

\graphicspath{{./}{figures/}}

\begin{document}

\title{Detection of p-mode Oscillations in HD 35833 with NEID and TESS}

\correspondingauthor{Arvind F.\ Gupta}
\email{arvind@psu.edu}

\author[0000-0002-5463-9980]{Arvind F.\ Gupta}
\affil{\PSUAA}
\affil{\PSUCEHW}

\author[0000-0002-4927-9925]{Jacob Luhn}
\affil{\UCI}

\author[0000-0001-6160-5888]{Jason T.\ Wright}
\altaffiliation{NEID Project Scientist}
\affil{\PSUAA}
\affil{\PSUCEHW}
\affil{\PSETI}

\author[0000-0001-9596-7983]{Suvrath Mahadevan}
\altaffiliation{NEID Principal Investigator}
\affil{\PSUAA}
\affil{\PSUCEHW}

\author[0000-0001-6545-639X]{Eric B.\ Ford}
\affil{\PSUAA}
\affil{\PSUCEHW}
\affil{\PSUICS}
\affil{\PSUCASt}

\author[0000-0001-7409-5688]{Gu{\dh}mundur Stef\'ansson}
\altaffiliation{\RUSSELL}
\affil{\Princeton}

%%% Begin Alphabetical List

\author[0000-0003-4384-7220]{Chad F.\ Bender}
\affil{\UA}

\author[0000-0002-6096-1749]{Cullen H.\ Blake}
\affil{\Penn}

\author[0000-0003-1312-9391]{Samuel Halverson}
\affil{\JPL}

\author[0000-0002-1664-3102]{Fred Hearty}
\affil{\PSUAA}
\affil{\PSUCEHW}

\author[0000-0001-8401-4300]{Shubham Kanodia}
\affil{\Carnegie}
\affil{\PSUAA}
\affil{\PSUCEHW}
\affil{\PSETI}

\author[0000-0002-9632-9382]{Sarah E.\ Logsdon}
\affil{\NOAO}

\author[0000-0003-0241-8956]{Michael W.\ McElwain}
\affil{\GoddardESAL} 

\author[0000-0001-8720-5612]{Joe P.\ Ninan}
\affil{\TIFR}

\author[0000-0003-0149-9678]{Paul Robertson}
\altaffiliation{NEID Project Scientist}
\affil{\UCI}

\author[0000-0001-8127-5775]{Arpita Roy}
\affil{\STScI}
\affil{\JHU}

\author[0000-0002-4046-987X]{Christian Schwab}
\affil{\Macquarie}

\author[0000-0002-4788-8858]{Ryan C. Terrien}
\affil{\Carleton}

%Chad Bender (confirmed)
%Cullen Blake (confirmed)
%Samuel Halverson (confirmed)
%Fred Hearty
%Shubham Kanodia (confirmed)
%Sarah Logsdon (confirmed)
%Michael McElwain (confirmed)
%Joe Ninan (confirmed)
%Paul Robertson (confirmed)
%Arpita Roy (confirmed)
%Christian Schwab (confirmed)
%Ryan Terrien (confirmed)

%\author{NEID Team Members}

\begin{abstract}

We report the results of observations of p-mode oscillations in the G0 subgiant star HD 35833 in both radial velocities and photometry with NEID and TESS, respectively. 
We achieve separate, robust detections of the oscillation signal with both instruments (radial velocity amplitude $A_{\rm RV}=1.11\pm0.09$ m s$^{-1}$, photometric amplitude $A_{\rm phot}=6.42\pm0.60$ ppm, frequency of maximum power $\nu_{\rm max} = 595.71\pm17.28$ $\upmu$Hz, and mode spacing $\Delta \nu = 36.65\pm0.96$ $\upmu$Hz) as well as a non-detection in a TESS sector concurrent with the NEID observations.
These data shed light on our ability to mitigate the correlated noise impact of oscillations with radial velocities alone, and on the robustness of commonly used asteroseismic scaling relations.
The NEID data are used to validate models for the attenuation of oscillation signals for exposure times $t<\nu_{\rm max}^{-1}$, and we compare our results to predictions from theoretical scaling relations and find that the observed amplitudes are weaker than expected by $>4\sigma$, hinting at gaps in the underlying physical models. %or star-to-star variability not captured by scaling relations.

\end{abstract}
%% Keywords should appear after the \end{abstract} command. 
%% See the online documentation for the full list of available subject
%% keywords and the rules for their use.

\section{Introduction}\label{sec:introduction}

The current generation of extreme precision radial velocity (EPRV) spectrographs, which includes NEID \citep{Schwab2016}, EXPRES \citep{Jurgenson2016}, ESPRESSO \citep{Pepe2021}, and MAROON-X \citep{Seifahrt2018}, is pushing past the 1 m s$^{-1}$ instrumental noise barrier and paving the way for the detection of 10 \cms{}  RV signals induced by Earth-mass exoplanets orbiting Sun-like stars.
While we may not achieve this feat with these instruments, they will afford us the opportunity to undertake more detailed studies of one of the largest remaining hurdles in the pursuit of Earth-analog exoplanet detection with radial velocities: intrinsic stellar variability \citep{Fischer2016,Crass2021}.
For even the quietest Sun-like stars, stellar variability manifests at the 1~m~s$^{-1}$ level and can easily mask the signals we seek to extract.
Without an appropriate means of accounting for these stellar signals, we will fail to detect and characterize sub-m s$^{-1}$ signals from low-mass, long-period exoplanets in spite of the advantage conferred by recent advances in instrumentation.

Efforts to mitigate the impact of stellar RV variations on exoplanet detection have found success by leveraging our understanding (albeit incomplete) of the underlying physical processes to model these signals rather than na\"ively treating them as white noise.
It has become commonplace, for example, to use Gaussian processes (GPs) to model stellar activity signals and disentangle them from exoplanet-induced Doppler shifts \citep[e.g.,][]{Haywood2014,Rajpaul2015,Foreman-Mackey2017,Gilbertson2020,Langellier2021,Zhao2022}.
But this strategy is only applicable when it is feasible to obtain many observations on the timescale over which the stellar signals remain correlated.
Short-timescale variations, such as acoustic, pressure-driven ``p-mode'' oscillations, necessitate a different approach. For Sun-like main sequence stars, p-mode oscillations produce RV signals that vary with periods of 5--10 minutes.
\citet{Chaplin2019} outline a detailed model for how the duration of an exposure affects the residual amplitude of RV signals due to stellar p-mode oscillations, showing that in principle, one can filter oscillation signals to the $<10$~\cms{} level with exposure times typical of current EPRV surveys.
Using an 8-hour $\alpha$ Cen A RV time series collected by \citet{Butler2004} with the UVES instrument on the VLT, 
\citet{Chaplin2019} show that the predicted residual amplitude does indeed match the observed RV signal for exposure times longer than typical oscillation periods, validating the general behavior of their model.
This model has been used to guide observing strategies for EPRV exoplanet searches \citep[e.g.,][]{Blackman2020,Gupta2021} and to inform simulations of future surveys \citep{Luhn2022}. \citet{Luhn2022} present a more comprehensive picture of the RV noise contributed by oscillations with a model that describes not only the impact on individual observations, but also correlations across multiple observations.
\rrev{
When applying these models to stars other than the Sun, for which we often lack empirical measurements of the oscillation frequencies and amplitudes, we typically rely on scaling relations to predict the asteroseismic parameters from known stellar properties \citep[e.g.,][]{Kjeldsen1995}. However, these scaling relations are imperfect, as they do not capture the full suite of properties that affect the oscillation signal.}
If this approach is to be effective on large samples of stars, we must therefore pursue a more rigorous understanding of p-mode oscillations and the radial velocity signals they induce.

In this work, we present the results of both simultaneous and asynchronous observations of p-mode oscillations in the subgiant star HD 35833 in radial velocities and photometry with NEID and \rrev{the Transiting Exoplanet Survey Satellite \citep[TESS;][]{Ricker2015}}, respectively. We describe the observations and data reduction in Section \ref{sec:observations}. Asteroseismic analyses of these data, detailed in Section \ref{sec:astero}, yield detections of suppressed\footnote{\rrevb{In this work, we use the term ``suppress'' to refer to a partial reduction in oscillation amplitude rather than a total absence of power.}} oscillation signals with NEID and one sector of TESS data, as well as a non-detection in a different TESS sector. The results shed light on our ability to mitigate the correlated noise impact of oscillations with radial velocities alone, and on the accuracy of stellar parameter scaling relations in predicting the frequencies and amplitudes of p-mode oscillations. In Section \ref{sec:discussion}, we discuss potential mechanisms for the observed oscillation amplitude suppression as well as implications of our results for future RV exoplanet searches and follow-up observations.

\section{Observations}\label{sec:observations}

\subsection{TESS Photometry}

\begin{figure*} 
    \centering
    \includegraphics[width=1.02\linewidth]{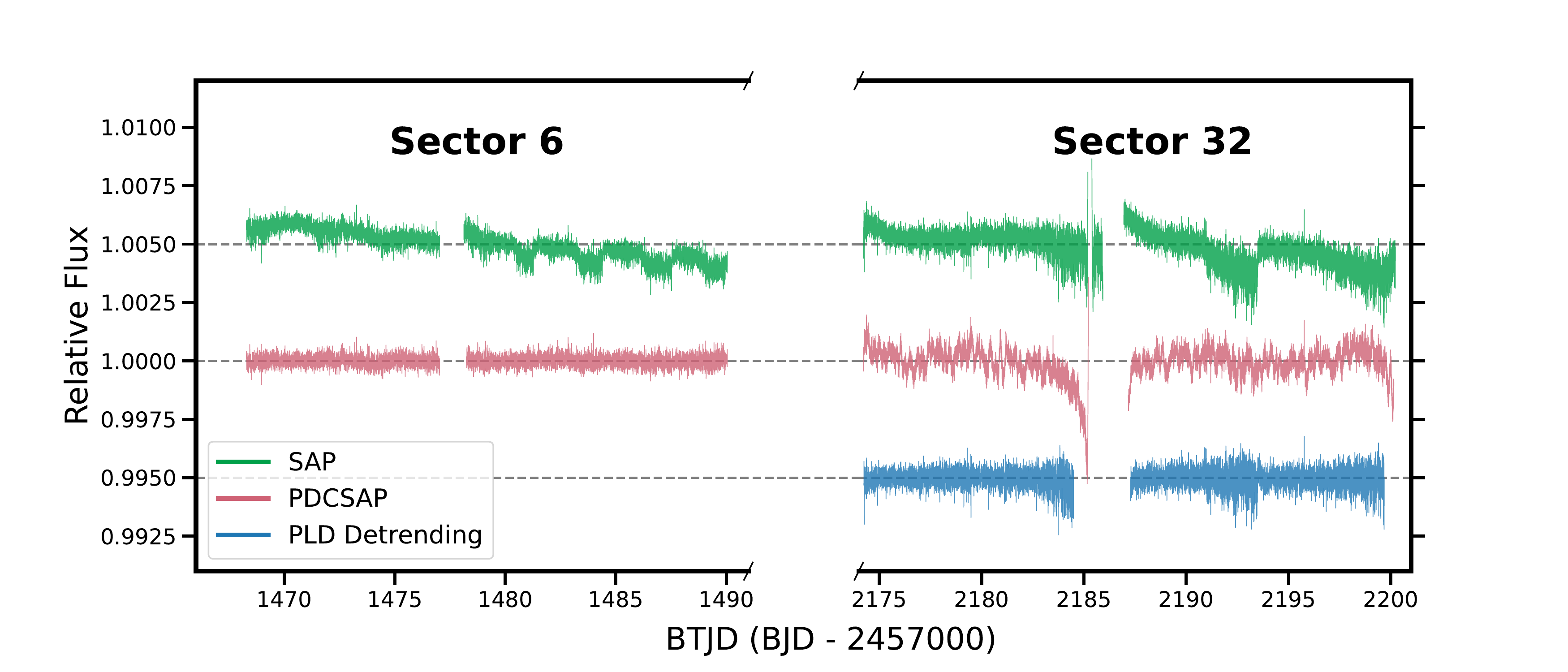}
    \caption{Cycles 1 and 3 TESS light curves for HD 35833. We show both the Sector 6 (left) and Sector 32 (right) 2-minute cadence data. From top to bottom: SAP (green), PDCSAP (pink), and custom PLD-deterended (blue; Sector 32 only). The photometric analysis herein makes use of the PDCSAP reduction (RMS CDPP $\approx 60$ ppm) for Sector 6, and the PLD-detrended reduction (RMS CDPP $\approx 100$ ppm) for Sector 32.}
    \label{fig:tess_lc}
\end{figure*}

\begin{figure*} 
    \centering
    \includegraphics[width=1.02\linewidth]{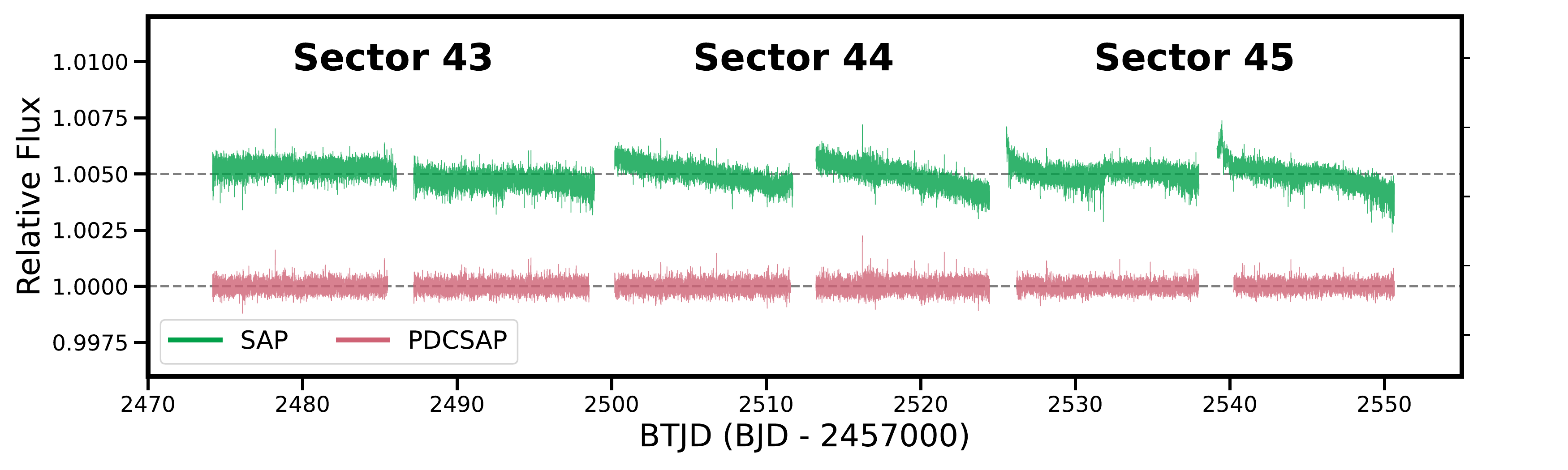}
    \caption{Cycle 4 TESS light curve for HD 35833. We show both the 2-minute cadence data for Sectors 43, 44, and 45. From top to bottom: SAP (green) and  PDCSAP (pink). The photometric analysis herein makes use of the PDCSAP reduction (RMS CDPP $\approx 60$ ppm) for the combined light curve.}
    \label{fig:tess_lc2}
\end{figure*}

TESS observed the subgiant star HD 35833 (HIP 25589, TIC 302423299) during Cycles 1 (Sector 6; 2018 December 11 - 2019 January 7), 3 (Sector 32; 2020 November 19 - 2020 December 17), and 4 (Sectors 43, 44, and 45; 2021 September 16 - 2021 December 2) at a 2-minute cadence. We obtain photometric data for all five sectors from the TESS Science Processing Operations Center \citep[SPOC;][]{Jenkins2016}, and we show the simple aperture photometry (SAP) and pre-search data conditioned simple aperture photometry (PDCSAP) light curves in Figures \ref{fig:tess_lc} and \ref{fig:tess_lc2}. The PDCSAP module \citep{Stumpe2012,Smith2012,Stumpe2014} is designed to correct for long term instrument systematics, background trends, and other sources of noise while preserving astrophysical signals on shorter timescales. For Sectors 6 and 43-45, we use the PDCSAP reduction for all photometric analysis herein. For Sector 32, however, we see unexpected, high-frequency flux variations in the PDCSAP reduction with amplitudes $>500$ ppm, nearly an order of magnitude greater than the $\sim$60 ppm combined differential photometric precision (CDPP) observed in the other sectors. The absence of this structure in the SAP light curve for Sector 32 suggests the signal is not astrophysical in nature but rather an artifact introduced by the pipeline detrending method.

\begin{figure}
    \centering
    \includegraphics[width=1.1\linewidth]{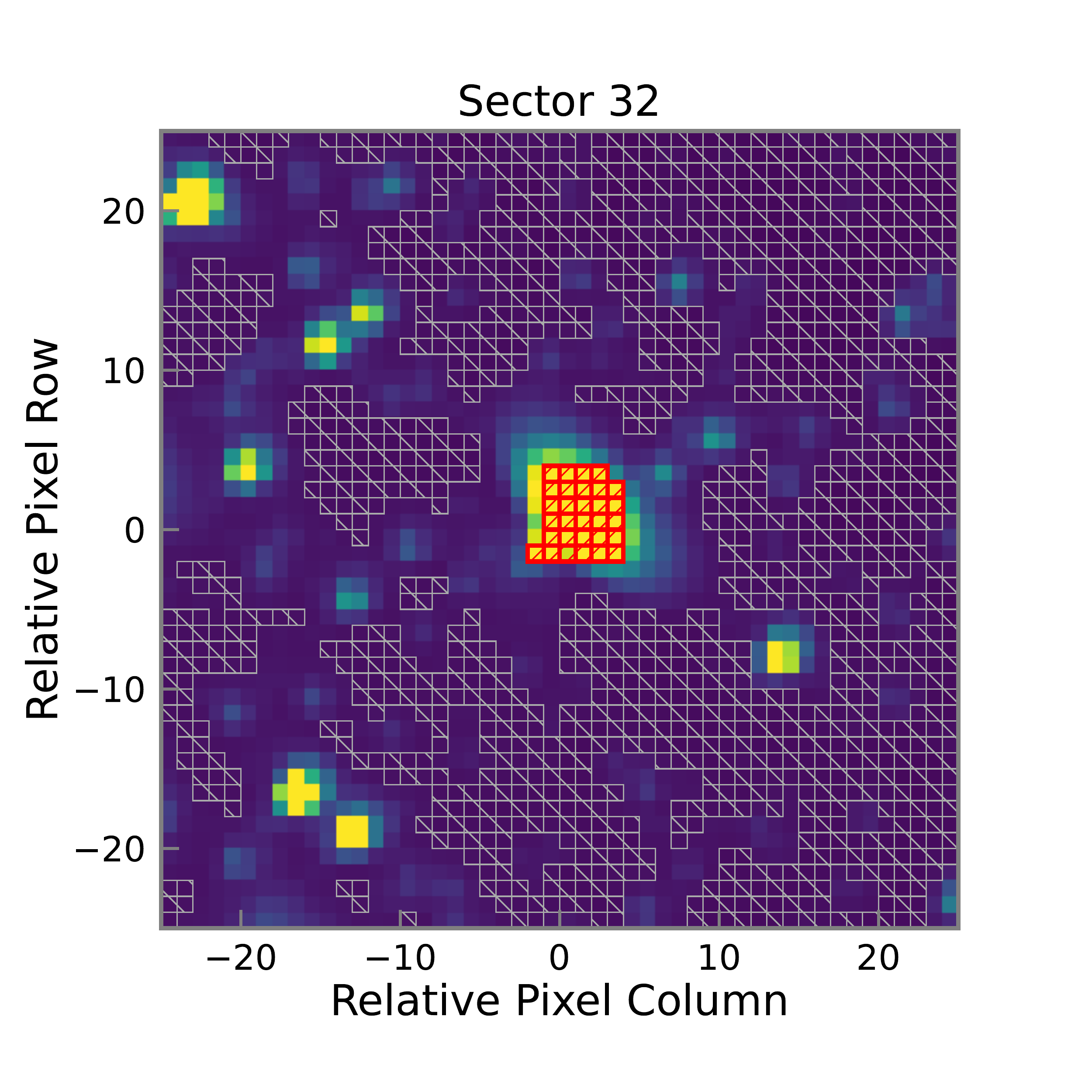}
    \caption{TESS FFI cutout used for PLD detrending of the HD 35833 Sector 32 light curve. The stellar aperture is shown in red and the background aperture is shown in grey.}
    \label{fig:tess_s32_cutout}
\end{figure}

We re-process the Sector 32 TESS photometry using the Pixel Level Decorrelation (PLD) toolkit in \texttt{lightkurve} \citep{LightkurveCollaboration2018}. This method, described in detail by \citet{Deming2015} and \citet{Luger2016,Luger2018}, uses information from nearby pixels to construct a model for the instrumental noise in the region of interest, which can then be subtracted from the pixels in the stellar aperture to produce a systematics-corrected light curve. 
We extract a 50 pixel $\times$ 50 pixel cutout from the 10-minute cadence full-frame image (FFI) data centered on HD 35833 (Figure \ref{fig:tess_s32_cutout}), and we calculate the noise model using the \texttt{lightkurve.PLDCorrector.correct} method. We restrict the corrector model to 5 PCA components so as not to overfit the data, though we note that the result is relatively insensitive to the number of components for $\leq10$ components except at the edges of the light curve.  For this correction, we use the default generated background aperture but we manually fix the stellar aperture to be the same as the pipeline aperture from the 2-minute cadence data. There are several known sources in this aperture, but the brightest of these is nearly 7 magnitudes fainter than the target in the \textit{Gaia} passband (Gaia DR3 3391121978660964992, $G=13.601$) and the contamination ratio is just $0.00335$. We do not expect blending to be a significant source of uncertainty.
We interpolate the noise model onto the 2-minute cadence time stamps and subtract it from the SAP flux. Finally, a cubic polynomial is fit to the result to remove any remaining long-term trends.

We also explored detrending the Sector 32 light curve using cotrending basis vectors (CBV)  following the methods of \citet{Lund2021}. This method did not produce a cleanly detrended light curve for HD 35833, however, so we use the final PLD-corrected light curve (Figure \ref{fig:tess_lc}; CDPP $\approx100$ ppm) for our analysis of the Sector 32 photometry.

%%%%%%%
%------
%%%%%%%

\subsection{NEID Radial Velocity Observations}

We observed HD 35833 with the NEID spectrograph \citep{Schwab2016,Robertson2019} on the WIYN\footnote{The WIYN Observatory is a joint facility of the NSF's National Optical-Infrared Astronomy Research Laboratory, Indiana University, the University of Wisconsin-Madison, Pennsylvania State University, the University of Missouri, the University of California-Irvine, and Purdue University.} 3.5m telescope at Kitt Peak National Observatory on the night of 2020 December 12 UT. We observed the star for $5.5$ hours at a 2-minute cadence (90 second exposures), following the target form near-zenith to airmass = $2.0$. While the intent was to obtain an 8-hour baseline, the observing window was limited by poor weather conditions during the first half of the night. The 2-minute cadence was selected so that we could achieve a high radial velocity precision for each exposure while also finely sampling the stellar oscillations. We observed using the NEID high resolution (R$\sim 113,000$) mode, taking simultaneous etalon calibration frames with the \rrev{optical density (OD)} 1.3 neutral density filter. During the observing sequence, we obtained 166 spectra with a median per-resolution-element S/N of 76 at 550 nm. The S/N of the exposures varies as a result of transparency changes due to intermittent cloud cover, but there is no trend with changing airmass.

%%%%%%%
%------
%%%%%%%

\begin{figure}
    \centering
    \includegraphics[width=1.05\linewidth]{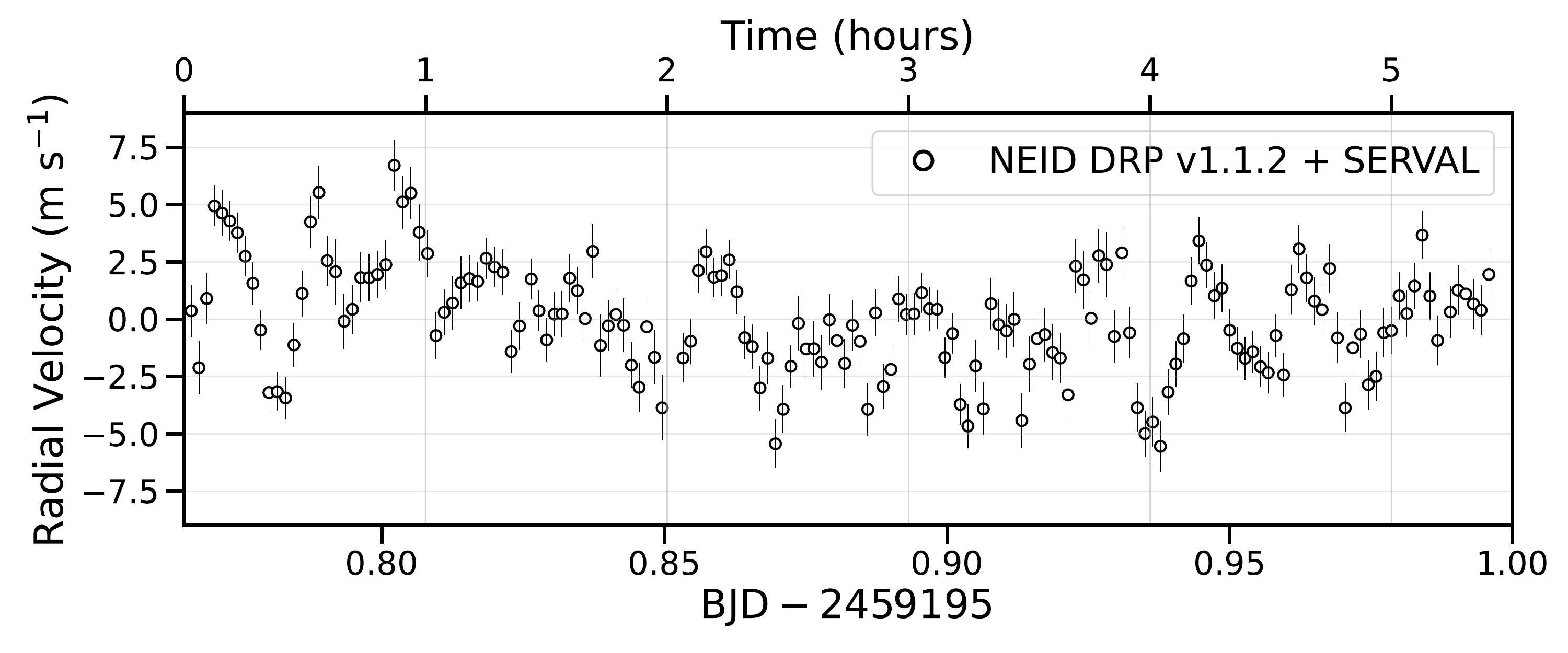}
    \caption{NEID RV measurements for HD 35833. RVs are calculated using a modified version of the SERVAL template-matching algorithm applied to 1D extracted spectra from the NEID DRP, with low S/N outliers removed.}
    \label{fig:raw_rvs}
\end{figure}

The NEID data were processed with version $1.1.2$ of the NEID Data Reduction Pipeline\footnote{\url{https://neid.ipac.caltech.edu/docs/NEID-DRP/}} (DRP).
We remove two outliers with S/N $<50$ leaving 164 high-quality spectra.
We consider two sources of RVs for our analysis: data products taken directly from the NEID DRP, which uses the Cross-Correlation Function \citep[CCF;][]{Baranne1996} method to calculate RVs, and an independent derivation using the template-matching method with a modified version of the \texttt{SpEctrum Radial Velocity AnaLyser} \citep[\texttt{SERVAL}][]{Zechmeister2018} analysis code, optimized for use for NEID spectra as described in \cite{Stefansson2022}. For the \texttt{SERVAL} reduction, we extracted RVs using order indices 69-153 spanning the wavelength region from 398nm to 895nm.
The two RV streams are fully consistent with each other, but we elect to use the \texttt{SERVAL} RVs for the remainder of this work as this method yields a better median RV precision (1.05 m s$^{-1}$) than the CCF-derived RVs from the DRP (1.70 m s$^{-1}$). The \texttt{SERVAL} radial velocity time series for HD 35833 is shown in Figure \ref{fig:raw_rvs}.

\section{Stellar Parameters}

The stellar parameters used in this work are taken from the Final Luminosity Age Mass Estimator (FLAME) module of the \textit{Gaia} Data Release 3 Astrophysical parameters inference system (Apsis) pipeline \citep{Creevey2022}.  
For bright stars, the FLAME module processes separate sets of inputs, from each of the GSP-Phot and GSP-Spec modules, producing two corresponding sets of stellar parameters \citep{Creevey2022}. We choose to use the FLAME/GSP-Phot results rather than the FLAME/GSP-Spec results for this work, as \citet{Fouesneau2022} and \citet{Recio-Blanco2022} note significant systematic biases in the $\log g$ values reported by the GSP-Spec module.
FLAME provides values for $R_\star$, $M_\star$, and $L_\star$, from which we also calculate $\teff$ and $\log g$. These parameters and their associated uncertainties are listed in Table \startable.

\begin{deluxetable*}{lccc}
\tablecaption{Summary of Stellar Parameters for HD 35833 \label{tab:stellar_params}}
\tablehead{\colhead{~~~Parameter}&  \colhead{Value}&
\colhead{Description}&
\colhead{Reference}}
\startdata
\multicolumn{4}{l}{\hspace{-0.2cm}Alternate Identifiers:}  \\
~~~TIC & 302423299 & TESS Input Catalog & Stassun \\
~~~Gaia DR3 &  3391121978660968576 & Gaia & Gaia DR3\\
~~~HIP & 25589 & HIPPARCOS Catalog  &  HIPPARCOS \\
\multicolumn{4}{l}{\hspace{-0.2cm} Coordinates and Parallax:} \\
~~~$\alpha_{\mathrm{J2016}}$ &    05:28:09.20 & Right Ascension (RA) & Gaia DR3\\
~~~$\delta_{\mathrm{J2016}}$ &   +16:26:19.61 & Declination (Dec) & Gaia DR3\\
~~~$\varpi$  & $14.89 \pm 0.02$ &  Parallax (mas) & Gaia DR3 \\
\multicolumn{4}{l}{\hspace{-0.2cm} Broadband photometry:}  \\
~~~TESS &  $6.2426 \pm 0.0061$ & TESS & Stassun \\
~~~$G$  &  $6.6975 \pm 0.0001$ & Gaia & Gaia DR3\\
~~~$B_p$ & $7.0194 \pm 0.0004$ & Gaia & Gaia DR3\\
~~~$R_p$  &  $6.2005 \pm 0.0005$ & Gaia & Gaia DR3 \\
\multicolumn{4}{l}{\hspace{-0.2cm} Derived Stellar Parameters:}\\
~~~$\teff$ & $5684.2\pm64.7$ & Effective Temperature (K) &  Gaia DR3\\
~~~$L_\star$ & $7.147\pm0.037$ &  Luminosity (${\rm L}_\odot$) & Gaia DR3\\
~~~$R_\star$ & $2.756\pm0.056$ & Stellar Radius (${\rm R}_\odot$)   &  Gaia DR3\\
~~~$M_\star$ & $1.423\pm0.041$ & Stellar Mass (${\rm M}_\odot$)  &  Gaia DR3\\
~~~$\log g$ & $3.710\pm0.030$ & Surface Gravity ($\log$ (g cm$^{-3}$)) &  Gaia DR3\\
\enddata
\tablenotetext{}{References are: HIPPAROCS \citep{ESA1997}, Stassun \citep{Stassun2019}, Gaia DR3 \citep{GaiaCollaboration2022}}

\end{deluxetable*}

\section{Asteroseismic Analysis}\label{sec:astero}
\subsection{Asteroseismic scaling relations for p-mode oscillations}\label{sec:astero_scaling}

Several asteroseismic quantities, including the frequencies, amplitudes, and frequency spacing of p-mode oscillations, are expected to scale with fundamental stellar properties. \citet{Kjeldsen1995} present a set of theoretically motivated and empirically validated scaling relations for these quantities, which we use here to predict the parameters of the oscillation signals that we observe with TESS and NEID.

The frequency of maximum power for the oscillation power excess scales as 
\begin{equation}\label{eq:numax}
    \nu_{\rm max} = \upnu_{\rm max, \odot} \frac{g/{\rm g}_\odot}{\sqrt{T_{\rm eff}/{\rm T}_{\rm eff, \odot}}}
\end{equation}
and $\Delta\nu$, the large oscillation mode spacing, is given by
\begin{equation}\label{eq:deltanu}
    \Delta\nu = \Delta\upnu_\odot \sqrt{\rho_\star/\uprho_\odot}
\end{equation}
where the above scaling relations are derived by \citet{Kjeldsen1995} and \rrev{we assume $\upnu_{\rm max, \odot} = 3090 \upmu$Hz and  $\Delta\upnu_\odot = 135.1 \upmu$Hz for the solar values \citep{Campante2016}}. Based on the stellar parameters given in Table \startable, we calculate the predicted values of $\nu_{\rm max}$ and $\Delta\nu$ for HD 35833 to be  $580.13\pm28.91\upmu$H\rrevb{z} and  $35.22\pm2.36\upmu$HZ, respectively. These values, along with other derived asteroseismic quantities, are shown in Table \asterotable. The quoted uncertainties are propagated from the uncertainties on the stellar parameters, and we do not account for intrinsic scatter in the asteroseismic scaling relations.

\begin{deluxetable*}{lccccc}
\tablecaption{Predicted and Observed Asteroseismic Parameters for HD 35833 \label{tab:astero_params}}
\tablehead{\colhead{~~~Parameter}&  \colhead{Scaling Relations}&  \colhead{TESS Sector 6} &
\colhead{TESS Sector 32}&
\colhead{TESS Sectors 43/44/45}&
\colhead{NEID}}
\startdata
~~~$\nu_{\rm max}$ ($\upmu$Hz) &  $580.13\pm28.91$ & $582.54\pm65.42$ & \textemdash& $595.71\pm17.28$ & \textemdash\\
~~~$\Delta\nu$  ($\upmu$Hz)& $35.22\pm2.36$ & $33.80\pm1.17$ & \textemdash & $36.65\pm0.96$& \textemdash\\
~~~$A_{\rm phot}$ (ppm) & $7.42\pm0.27$ & $5.65\pm1.25$ & $<10.90$ $(99\%)$& $6.42\pm0.60$ & \textemdash\\
                        & &  & $<7.89$ $(50\%)$ & & \\
~~~$A_{\rm RV}$ (m s$^{-1}$) &  $2.86\pm0.10$ & \textemdash& \textemdash& & $1.11\pm 0.09$ \\
%\hline
%~~~White Noise & & & & & \\
%~~~(ppm$^2/\upmu$Hz) &  $2.86\pm0.10$ & \textemdash& \textemdash& & $1.11\pm 0.09$ \\
\enddata
\end{deluxetable*}

The expected amplitudes of the oscillation signals observed with both TESS and NEID can also be calculated using scaling relations from \citet{Kjeldsen1995}. 
For TESS, the expected amplitude of radial, $l=0$, oscillation modes follows
\begin{equation}\label{eq:aphot}
    A_{\rm phot}  ={\rm A}_{\rm phot,\odot}
    \beta\left(\frac{L_\star}{{\rm L}_\odot}\right)^s\left(\frac{{\rm M}_\odot}{M_\star}\right)^s\left(\frac{\teff, \star}{{\rm T}_{\rm eff, \odot}}\right)^{-2}
\end{equation}
where we assume ${\rm A}_{\rm phot,\odot}=2.125$ ppm for the TESS bandpass, as calculated by \citet{Campante2016} based on a derivation of the amplitude in the \textit{Kepler} bandpass by \citet{Ballot2011}.
We also apply the correction factor $\beta$, first introduced by \citet{Chaplin2011beta}, to account for observed deviation from the above scaling relation for relatively hot stars:

\begin{equation}\label{eq:aphotbeta}
    \beta = 1 - \exp\left(-\frac{T_{\rm red}-\teff}{1550 {\rm\ K}}\right)
\end{equation}
\begin{equation}\label{eq:aphottred}
    T_{\rm red} = 8907(L_\star/{\rm L}_\odot)^{-0.093} {\rm\ K}.
\end{equation}
We note that this calculation disregards any dependence on stellar activity; this is discussed in further detail in Section \ref{sec:activity}.  If we assume $s=1$ for the mass and luminosity scaling in Equation \ref{eq:aphot}, as in \citet{Campante2016} and \citet{Chaplin2011beta}, then using the derived stellar parameters listed in Table \startable, we find $A_{\rm phot,predicted}=7.42\pm0.27$ ppm.

It bears emphasizing that Equation \ref{eq:aphot} is specifically for the amplitudes of the radial oscillation modes, which differ from $A_{\rm max,phot}$. $A_{\rm max,phot}$ is defined as the amplitude of a Gaussian fit to the oscillation power excess (with units ppm$^2$/$\upmu$Hz), and it is related to $A_{\rm phot}$ via the equation 
\begin{equation}\label{eq:radial_correction}
    A_{\rm phot} = \sqrt{\left(A_{\rm max, phot}\frac{\Delta\nu}{c}\right)}
\end{equation}
where the factor of $c$, the effective number of oscillation modes per spherical harmonic order, corrects for the contributions of non-radial ($l\neq0$) modes to the power excess.

For NEID, the expected radial velocity amplitude of the oscillation signal is given by
\begin{equation}\label{eq:amaxrv}
    A_{\rm max, RV} = {\rm A}_{\rm max, RV, \odot} \left(\frac{L_\star}{{\rm L}_\odot}\right)\left(\frac{{\rm M}_\odot}{M_\star}\right).
\end{equation}
We adopt ${\rm A}_{\rm max, RV, \odot}= 0.19$ m s$^{-1}$ from \citet{Chaplin2019}. The radial velocity amplitude for the $l=0$ modes is calculated as in Equation \ref{eq:radial_correction}, using $c=4.09$ as derived by \citet{Kjeldsen2008}. Here, we find $A_{\rm RV,predicted}=2.86\pm0.10$ m s$^{-1}$.
\rrev{We note that the correction factor $c$ depends on the relatively visibilities of different oscillation modes, which may differ for stars of different masses and evolutionary states. The dipolar $l=1$ mode in particular have been shown to be weaker in some cases for evolved stars \citep{Stello2016}, which may influence the accuracy of the predicted values of the oscillation amplitudes in this work. We comment on this possibility in Section \ref{sec:other_supp}.}

\subsection{Photometric Analysis}\label{sec:tess_astero}

\begin{figure}
    \centering
    \includegraphics[width=1.0\linewidth]{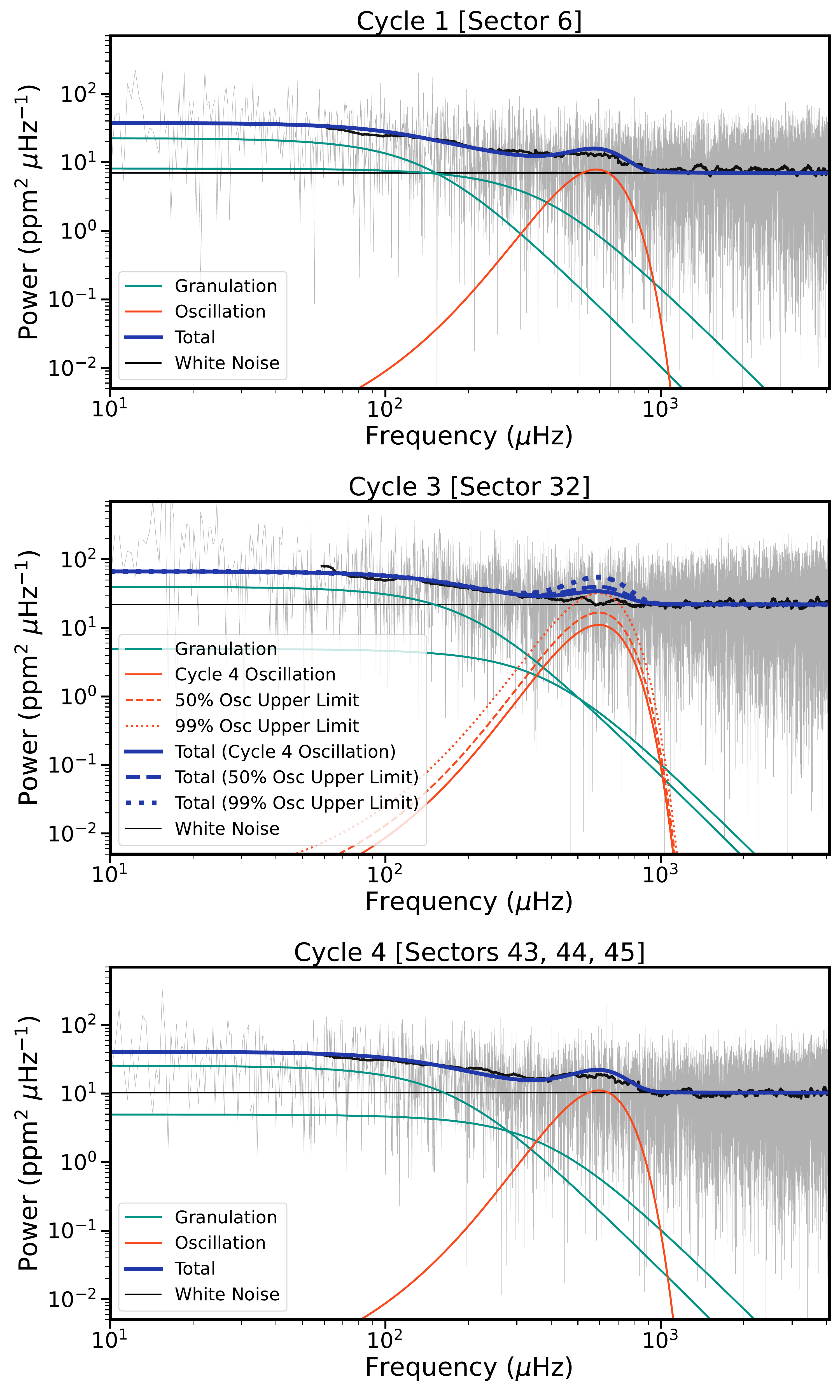}
    \caption{\rrev{Power spectral density for HD 35833 TESS Sector 6 (top), Sector 32 (middle), and the Cycle 4 data (bottom), with fine sampling in light grey and a boxcar-smoothed power spectrum in black. For Sectors 6 and 43-45, we show the individual oscillation (orange), granulation (green), and white noise (grey horizontal line) components as fit by \texttt{pySYD} as well as the sum of these components (blue). For Sector 32, we show the fitted granulation and white noise components, as well as injected oscillation signals with frequencies set by the signal detected in the Cycle 4 data and amplitudes scaled by $1.23\times$ (dashed line) and $1.70\times$ (dotted line) for the 50\% and 99\% detection limits, respectively.}}
    \label{fig:phot_psd}
\end{figure}

We conduct a global asteroseismic analysis of the TESS photometry using the \texttt{pySYD} fitting package \citep{Chontos2021}, treating each of the Cycle 1 (Sector 6), Cycle 3 (Sector 32) and Cycle 4 (Sectors 43-45) light curves as independent data sets.
\rrev{The resulting PSDs along with the fitted granulation and oscillation components are shown in Figure \ref{fig:phot_psd}.
We also include the white noise values ($9.59$ ppm$^2/\upmu$Hz, $21.59$ ppm$^2/\upmu$Hz, and $11.96$ ppm$^2/\upmu$Hz, for Cycles 1, 3, and 4, respectively), which are calculated as the median power in the range $2000\upmu$Hz to the Nyquist frequency of $4167\upmu$Hz.}
From the Sector 6 light curve, we obtain a significant detection of the oscillation signal with $\nu_{\rm max} = 582.54\pm65.42\upmu$Hz and large frequency spacing $\Delta\nu = 33.80\pm1.17 \upmu$Hz, both of which are consistent with the values predicted by theoretical scaling relations (Equations \ref{eq:numax} and \ref{eq:deltanu}). The amplitude of the Gaussian fit to the smoothed, background-corrected power excess is $A_{\rm max,phot} = 2.80\pm1.23$ppm$^2/\upmu$Hz. As in \citet{Huber2019}, we then follow the prescription of \citet{Kjeldsen2008} and \citet{Kjeldsen2005} to correct for the contributions of non-radial modes and calculate the mean amplitude of the radial, $l=0$, modes following Equation \ref{eq:radial_correction}. The value of $c$ used in Section \ref{sec:astero_scaling} is not applicable here, however, as the relative visibilities of radial and nonradial modes differ between photometry and radial velocities. We calculate $c=2.96$ for the TESS bandpass, centered on $786.5$ nm, by linearly interpolating between the 500 nm and 862 nm values derived by \citet{Kjeldsen2008}, and we find that $A_{\rm phot} = 5.65\pm1.25$ ppm for the Sector 6 light curve. We obtain similar results for the Cycle 4 data, with $\nu_{\rm max} = 595.71\pm17.28\upmu$Hz, $\Delta\nu = 36.65\pm0.96 \upmu$Hz, and $A_{\rm max,phot} = 3.33\pm0.62$ppm$^2/\upmu$Hz, and we calculate $A_{\rm phot} = 6.42\pm0.60$ ppm. While both data sets had similar photometric precision, the asteroseismic quantities derived from the Cycle 4 light curve are more precise as a consequence of the longer time baseline.

For Sector 32, we do not detect the oscillation power excess with \texttt{pySYD}, likely because of the relatively poor photometric precision of the data (CDPP$\approx 100$ ppm for Sector 32 vs. CDPP $\approx 60$ ppm for the other data) \rrev{and consequently the significantly larger background noise in the power spectrum}.
We place upper limits on the photometric oscillation amplitude in the Sector 32 light curve via a set of injection recovery tests using the properties of the oscillation signal extracted from the Cycle 4 data. The background-corrected oscillation power excess, with height scaled to simulate a change in $A_{\rm phot}$, is used to generate an oscillation time series which is then injected into the Sector 32 light curve. The modified light curve is re-analyzed using \texttt{pySYD} to try to recover the injected signal. We achieve a 50\% recovery rate for injected signals with $A_{\rm phot} = 7.89$ ppm ($1.23\times$ the amplitude detected in Cycle 4) and a 99\% recovery rate for injected signals with $A_{\rm phot}=10.90$ ppm ($1.70\times$ the amplitude detected in Cycle 4). These results, illustrated in Figure \ref{fig:phot_psd}, show that we should not expect to detect the oscillation signal in the Sector 32 data if the photometric oscillation amplitude were the same as in the other sectors. In addition, the predicted amplitude is just below the 50\% detection threshold, so this result is consistent with expectations from scaling relations.

Though the non-detection in Sector 32 is unsurprising, the observed values of $A_{\rm phot}$ from Sector 6 and Sectors 43-45 differ slightly ($<2\sigma$ discrepancy) from the predicted value. These low amplitudes suggest that the oscillation signal for HD 35833 may be suppressed\rrevb{, or weakened,} by some mechanism that is not accounted for in the Equation \ref{eq:aphot} scaling relation. We discuss this possibility in detail in Section \ref{sec:discussion}.

\subsection{Radial Velocity Analysis}\label{sec:neid_astero}

\begin{figure}
    \centering
    \includegraphics[width=1.05\linewidth]{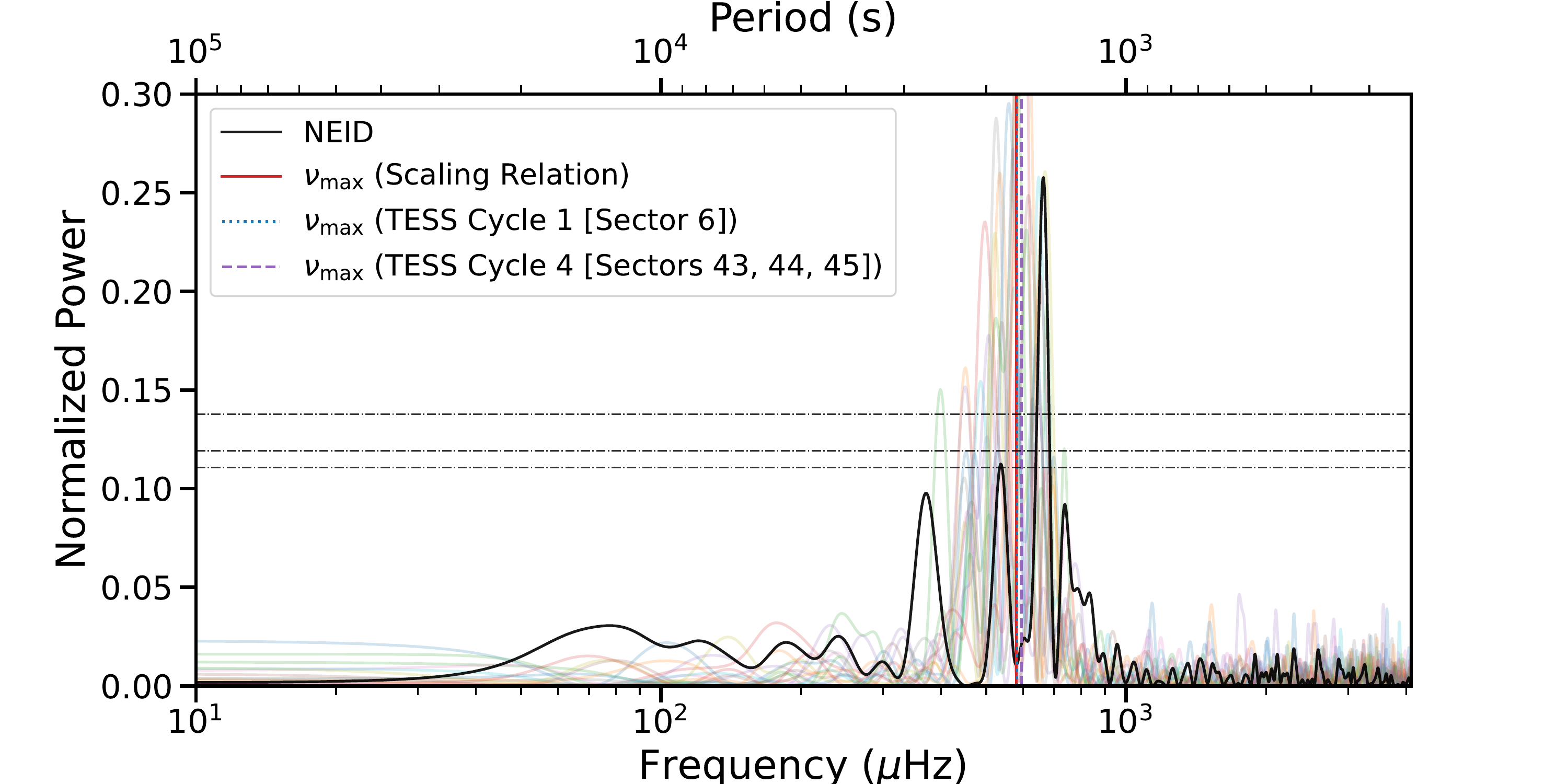}
    \caption{Lomb-Scargle periodogram of the detrended NEID observations of HD 35833. \rrev{We use the same frequency limits as in Figure \ref{fig:phot_psd} for ease of comparison.} We also show the frequency of maximum power for the oscillation signal as detected in TESS Cycles 1 (blue\rrevb{, dotted}) and 4 (purple\rrevb{, dashed}) and as predicted by scaling relations (red) as vertical lines, as well as the 10\%, 5\%, and 1\% false alarm levels (calculated via the \texttt{false_alarm_level()} method of the \texttt{astropy.timeseries.LombScargle} class) as horizontal lines \rrevb{(black, dot-dashed)}. For comparison, we show periodograms for a set of simulated oscillation time series with the same sampling and photon noise properties as the NEID observations (background colored lines).}
    \label{fig:rv_periodogram}
\end{figure}

The signals of interest for this work vary on timescales of less than an hour. Before proceeding with the RV analysis, we first fit and subtract a slowly varying quadratic trend from the data to remove any long timescale variation, e.g., supergranulation or uncorrected instrument drift. We then run a Lomb-Scargle periodogram analysis \citep{Lomb1976,Scargle1982} to search for the oscillation signal, and we detect several peaks near $\nu_{\rm max}$ as predicted by the scaling relation in Equation \ref{eq:numax} and as measured via the TESS data in Section \ref{sec:tess_astero} (Figure \ref{fig:rv_periodogram}). However, it is clear that the \rrev{\rstrike{individual modes are}periodogram is} not well resolved in frequency space due to the short baseline of the NEID observations\rrev{, so these data can not be used to measure the specific frequencies and amplitudes of individual modes. As we illustrate in Figure \ref{fig:rv_periodogram} via simulations of the oscillation time series, the observed periodogram structure can vary significantly from one $5.5$-hour realization to the next}. We therefore elect not to fit a Gaussian envelope to the oscillation power excess for the NEID data, and we instead analyze the RVs in the time domain.

%[Gaussian Process Decomposition \rnote{Jacob!}]
\subsubsection{Gaussian Processes Conditioning \& Decomposition}\label{sec:gp_decomp}
We perform a Gaussian process (GP) regression using the RV kernels for granulation and oscillations given in \citet{Luhn2022}, based on the photometric-to-RV scalings from \citet{Guo2022}. The kernel hyperparameters are set by $\teff$ and $\log g$ and as a result, the GP kernels are fully deterministic.

If the structure of the covariance matrix is known, a set of $n_1$ observations $(\mathbf{t_{1}},\mathbf{y_1})$ can be used to make predictions over a different sample  $(\mathbf{t_2},\mathbf{y_2})$ of length $n_2$, as they are jointly Gaussian
\begin{equation}
    \begin{bmatrix} \mathbf{y_1} \\ \mathbf{y_2} \end{bmatrix} \sim N\left(\begin{bmatrix} {\mu_1} \\ {\mu_2} \end{bmatrix}, \begin{bmatrix} \Sigma_{11} & \Sigma_{12} \\ \Sigma_{21} & \Sigma_{22}
    \end{bmatrix}
    \right).
\end{equation}

The resulting posterior distribution for $\mu_{2}$ \emph{conditioned} on the observations $(\mathbf{t_{1}},\mathbf{y_1})$ is then centered on the mean
\begin{equation}\label{eqn:mu2_single}
    \mu_{2|1} = \left( \Sigma_{11}^{-1} \Sigma_{12}\right)^{\top} \mathbf{y_1},
\end{equation}
with covariance
\begin{equation}
    \Sigma_{2|1} = \Sigma_{22} - \left(\Sigma_{11}^{-1}\Sigma_{12}\right)^{\top} \Sigma_{12},
\end{equation}
and standard deviation
\begin{equation}
    \sigma_{2|1} = \left[\mathrm{Diag}\left(\Sigma_{2|1}\right)\right]^{1/2}.
\end{equation}

The previous equations describe standard GP conditioning in the case of a single kernel. In our case, however, we have described the covariance matrix as the sum of two astrophysical GP kernels (granulation and oscillation), e.g.,
\begin{equation}
    \Sigma[i,j] = k_{a}(t_{i}, t_{j}) + k_{b}(t_{i},t_{j}).
\end{equation}
In this case, \autoref{eqn:mu2_single} describes the conditioned mean for the sum of two underlying processes. We wish to separate out the mean effect that each of these kernels contributes. We can write \autoref{eqn:mu2_single} more explicitly in this case as
\begin{equation}\label{eqn:mu2_sum}
    \mu_{2|1,a+b} = \left[ \Sigma_{11}^{-1} \left(\Sigma_{12,a}+\Sigma_{12,b}\right)\right]^{\top} \mathbf{y_1},
\end{equation}
If we define
\begin{align}\label{eqn:defs}
\mu_{2|1,a} &= \left(\Sigma_{11}^{-1} \Sigma_{12,a}\right)^{\top}\bf{y_1} \\
\mu_{2|1,b} &= \left(\Sigma_{11}^{-1} \Sigma_{12,b}\right)^{\top}\bf{y_1},
\end{align}
it is true that
\begin{equation}
     \mu_{2|1,a+b} = \mu_{2|1,a} + \mu_{2|1,b}
\end{equation}
where $\mu_{2|1,a}$ represents the conditioned time series described by the kernel $k_{a}(t_{i}, t_{j})$ and $\mu_{2|1,b}$ represents the conditioned time series described by the kernel $k_{b}(t_{i}, t_{j})$.
In this way, our conditioned mean $\mu_{2|1}$ can be broken up into its component GP kernels. We refer to this as GP \emph{decomposition} to isolate individual components of a multi-component GP kernel. While the decomposition of the conditioned \emph{means} is relatively trivial, the decomposition of the conditioned \emph{covariance} includes cross terms that cannot be assigned to an individual component. 

We wish to focus a moment on our definitions above for $\mu_{2|1,a}$ and $\mu_{2|1,b}$. It is important to note that these are not the same as conditioning the data solely using one kernel or the other, as \emph{both} kernels are still included in the construction of the covariance matrix $\Sigma_{11}$, which also includes the diagonal observation errors (i.e., photon noise). We can show this more explicitly by writing Equation \ref{eqn:mu2_single} in a slightly different way
\begin{equation}\label{eqn:mu2_single_rewrite}
    \mu_{2|1} =  \Sigma_{21} \Sigma_{11}^{-1} \mathbf{y_1}.
\end{equation}
The conditioned mean (\autoref{eqn:mu2_single_rewrite}) can be expanded to explicitly show each component of our granulation and oscillation GP sum
\begin{equation}
\begin{split}
    \mu_{2|1, gran+osc} =& \left(\Sigma_{21,gran}+\Sigma_{21,osc}\right) \times \\ & \left(\Sigma_{11,gran}+\Sigma_{11,osc}+\Sigma_{11,phot}\right)^{-1} \bf{y_1},
\end{split}
\end{equation}
which can be decomposed into a granulation component
\begin{equation}
\mu_{2|1, gran} = \left(\Sigma_{21,gran}\right)\left(\Sigma_{11,gran}+\Sigma_{11,osc}+\Sigma_{11,phot}\right)^{-1} \bf{y_1},
\end{equation}
and an oscillation component
\begin{equation}
\begin{split}
\mu_{2|1, osc} = & \left(\Sigma_{21,osc}\right) \times 
\\
& \left(\Sigma_{11,gran} +\Sigma_{11,osc}+\Sigma_{11,phot}\right)^{-1} \bf{y_1}.
\end{split}
\end{equation}

\begin{figure*}
    \centering
    \includegraphics[width=1.05\linewidth]{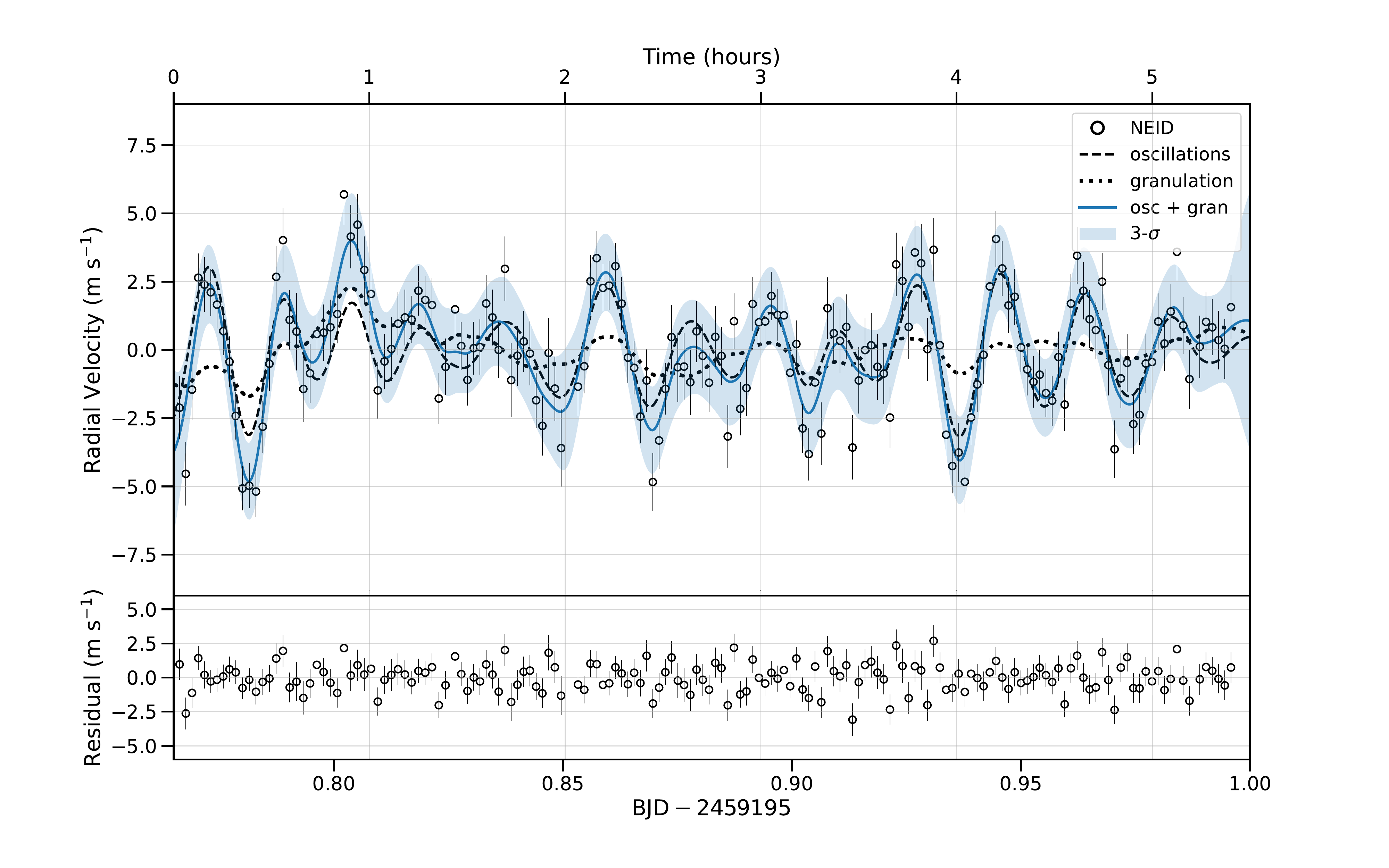}
    \caption{Detrended NEID RV time series for HD 35833 (top) and residuals (bottom). The individual oscillation and granulation signals, separated using the GP decomposition method described in Section \ref{sec:gp_decomp}, are shown as black dashed (oscillation) and dotted (granulation) lines. The sum of these signals is shown as a solid blue line. The residual signal is calculated by interpolating the oscillation and granulations signals to the observation time stamps and subtracting these from the detrended RVs.}
    \label{fig:decomp_rv_series}
\end{figure*}

We apply this GP conditioning and decomposition to the detrended NEID time series to isolate the individual oscillation and granulation components of the stellar radial velocity signal.
The resulting time series and the radial velocity residuals are shown in Figure \ref{fig:decomp_rv_series}.
%\rnote{[Probably need to add some more here about the actual decomposed time series]}

\subsubsection{RMS of decomposed time series}\label{sec:rms_breakdown}

One of the main conclusions of the \citet{Chaplin2019} study is that the residual amplitude of a stellar oscillation signal can be predicted if the stellar parameters are known, with appropriate caveats for stochastic mode excitation as discussed in Section \ref{sec:rv_amp}.
Because the residual amplitude manifests as an RMS when treating oscillations as a source of noise in a RV time series (e.g., for exoplanet surveys), this means that the oscillation contribution to the total RMS can be controlled by adjusting the integration time of one's observations.
With the isolated oscillation RV time series from the GP decomposition for HD 35833, we have the means to explore this prediction.

\begin{figure}
    \centering
    \includegraphics[width=1.0\linewidth]{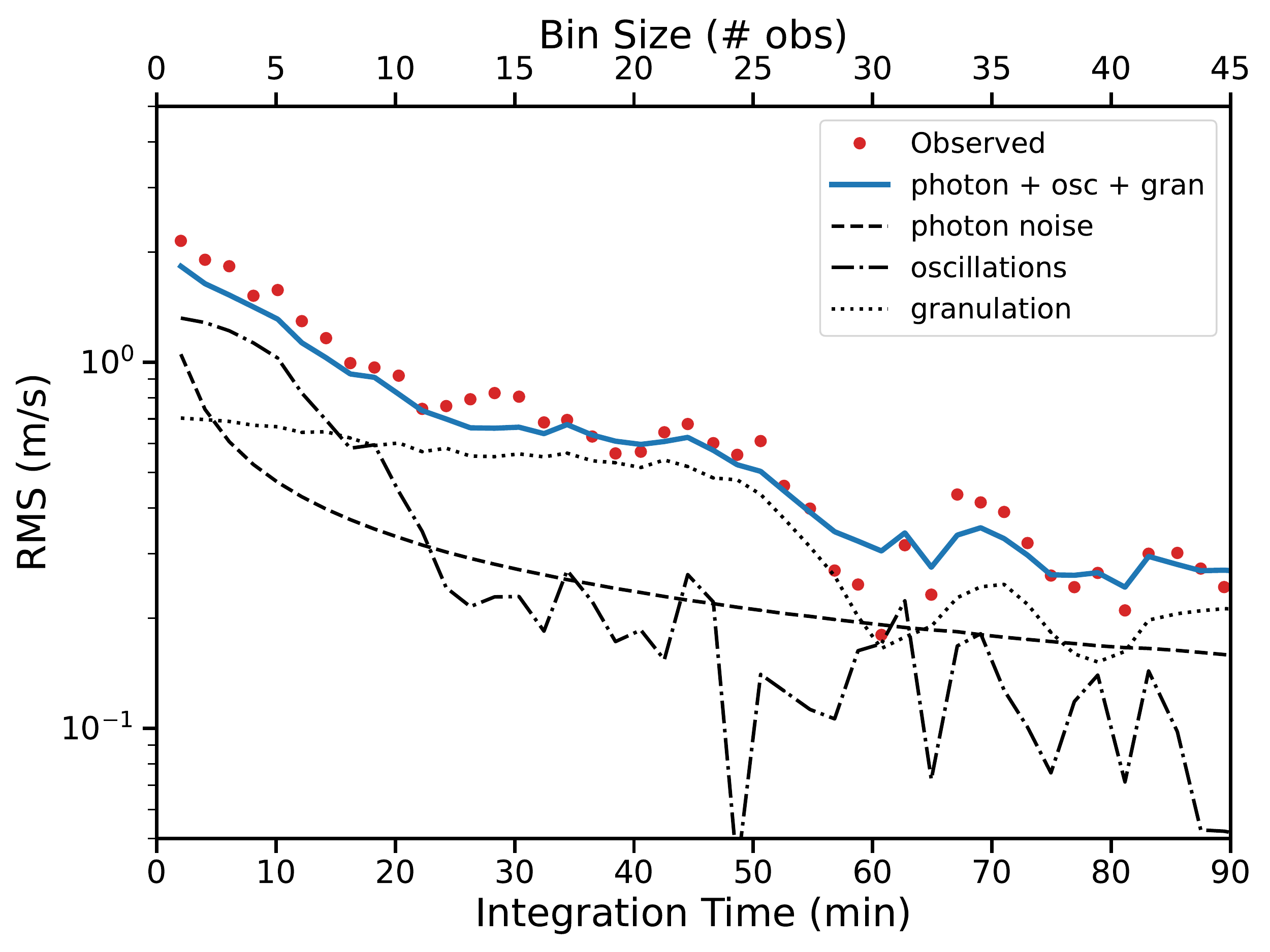}
    \caption{RV RMS as a function of integration time for NEID observations of HD 35833. We show the total observed RMS (red points), the photon noise as calculated from the information content of the spectra (black dashed line), the RMS of the decomposed oscillation (black dot-dashed line) and granulation (black dotted line) signals, and the sum of the three sources of RV ``noise'' (blue solid line). The shape of the oscillation signal is consistent with predictions from the \citet{Chaplin2019} model}.
    \label{fig:decomposed_rms}
\end{figure}

Here, we vary the integration time, $t_i$, by binning sets of consecutive exposures, so the integration time is defined as the sum of the on-sky exposure times, $t_e$, for all exposures in a bin \textit{and} inter-exposure readout times, $t_r$. For a bin with $M$ observations, then,
\begin{equation}
    t_i = Mt_e+(M-1)t_r.
\end{equation}
In Figure \ref{fig:decomposed_rms}, we show the RV RMS as a function of integration time for the total observed signal and each of the decomposed oscillation and granulation components. We also plot the measured photon noise uncertainties, as well as the sum of the three sources of ``noise'': photon noise, oscillations, \rrev{and} granulation.
We find that the GP decomposition does reliably capture the correlated nature of the oscillation signal, as the observed shape of the oscillation RMS curve agrees with predictions (Figure \ref{fig:rv_amp}). The total amplitude, however, is somewhat lower than expected, consistent with our analysis of the TESS photometry in Section \ref{sec:tess_astero} suggesting that the oscillation amplitudes are suppressed.

\subsubsection{Radial Velocity Oscillation Amplitudes}\label{sec:rv_amp}

\citet{Chaplin2019} lay out a detailed explanation of how one can predict the residual amplitudes of stellar p-mode oscillation signals for radial velocity observations given knowledge of the stellar properties.
In the \citet{Chaplin2019} model, oscillation amplitudes are attenuated by the factor 
\begin{equation}\label{eq:simple_transfer}
    \eta(\nu) = \sinc(\pi \nu t_e)
\end{equation}
where $\nu$ is the oscillation frequency and $t_e$ is the exposure time of the observation.
This transfer function is accurate when calculating the amplitude attenuation for a single continuous exposure, or, equivalently, a sequence of back-to-back, uninterrupted exposures. But in practice, most radial velocity instruments cannot achieve a 100\% duty cycle when taking multiple exposures; individual exposures will be separated by a non-zero readout time, during which the stellar oscillation signal is not being observed.
In this case, the transfer function must be scaled by a more complicated function of the exposure time, readout time, and total number of exposures, $N$
\begin{equation}\label{eq:gen_transfer}
    \eta(\nu)=\frac{1}{N}\sinc(\pi \nu t_e)\sum_{j=0}^N e^{\pi i\nu(2j-N+1)(t_e+t_r)}.
\end{equation}
We show the full derivation of this equation in the appendix.

\begin{figure}
    \centering
    \includegraphics[width=1.0\linewidth]{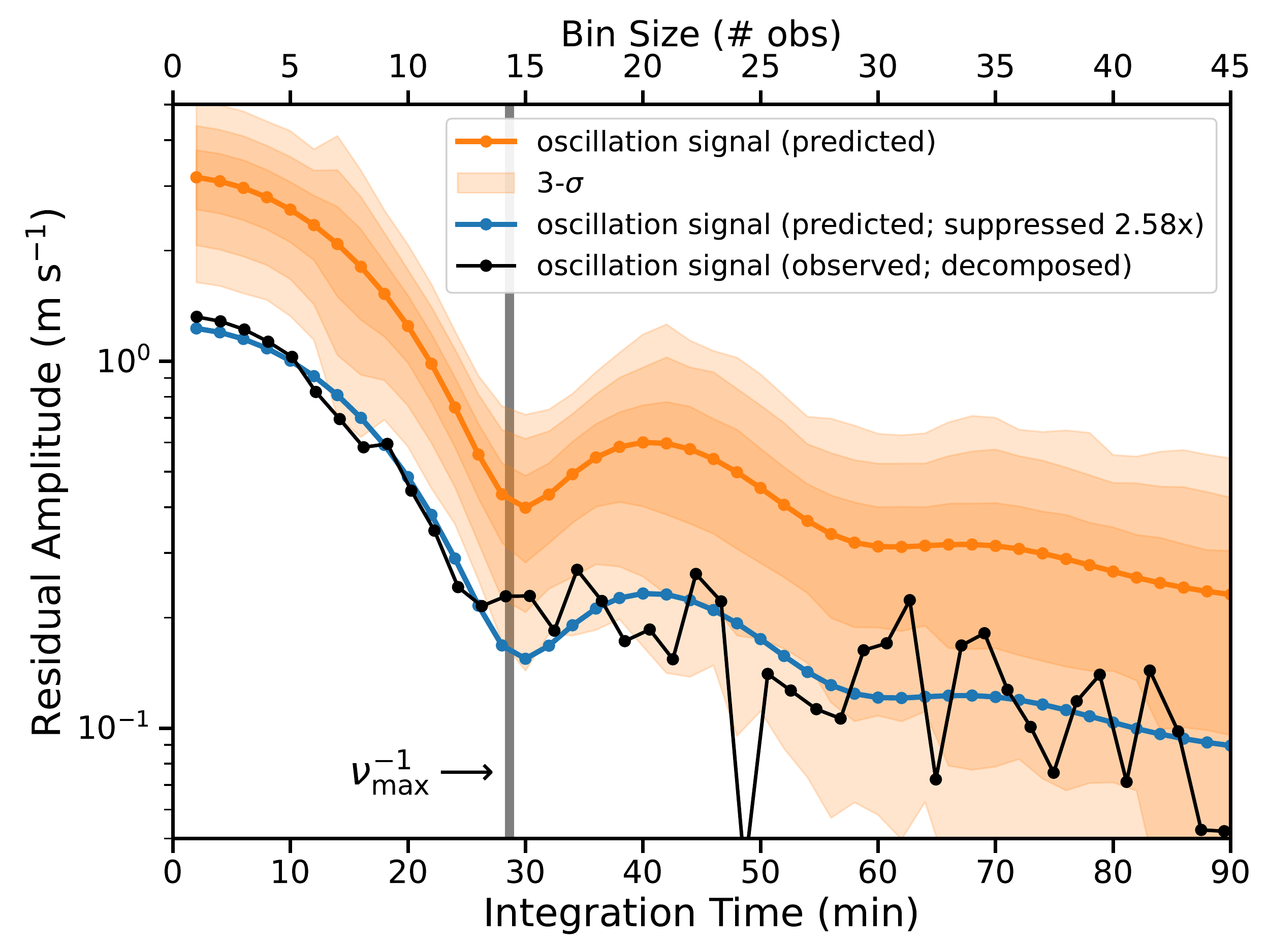}
    \caption{Predicted (orange) and observed, decomposed (black) residual amplitude for the decomposed p-mode oscillation signal in the NEID RVs for HD 35833. While the shapes are the same, the amplitude of the observed signal is suppressed relative to predictions by a factor of 2.58 (blue). We also indicate the predicted value of $\nu_{\rm max}^{-1}$ as a grey vertical line.}
    \label{fig:rv_amp}
\end{figure}

The predicted residual oscillation amplitude for our NEID observations, shown in Figure \ref{fig:rv_amp}, is calculated following the procedure outlined in \citet{Chaplin2019} but with the generalized form of the transfer function (Equation \ref{eq:gen_transfer}) used in place of the simple sinc function (Equation \ref{eq:simple_transfer}). 
Because the net oscillation signal is produced by many simultaneously beating modes that are stochastically excited and damped, this prediction only represents the median case. We expect the residual amplitude that we measure with NEID to deviate from this prediction somewhat, as it will depend on the exact interference pattern of the oscillations during the short, 5.5-hour observing window. To illustrate the expected distribution about the median case, we simulate an oscillation time series $10,000$ times, copying the NEID observing cadence and exposure times and randomizing the phases of the modes each time. We then bin the data and calculate residual amplitudes for each realization as described in Section \ref{sec:rms_breakdown}, and we compute the $16 - 84$, $2.3 - 97.7$, and $0.2 - 99.8$ percentile ranges to place $1\sigma$, $2\sigma$, and $3\sigma$ constraints on the residual amplitude we expect to observe in each bin.

As we show in Figure \ref{fig:rv_amp}, the residual amplitude of the observed oscillation signal is significantly lower than what we predict, falling outside of the $3\sigma$ contours for most integration times $\leq \nu_{\rm max}^{-1}$. As with the photometric data, these results suggest that the oscillations were suppressed. 

We fit for the \rrevb{reduction in amplitude} by taking the ratios of the predicted and observed residual amplitude curves in the range 2 minutes $\leq$ integration time $\leq$ 30 minutes; we do not expect the data to be sensitive to variations at the $<<1$ m s$^{-1}$ level even when binning dozens of observations. The mean of the ratios, weighted by the uncertainty on the oscillation RMS, is 2.58 in this range. That is, the observed residual amplitude is 2.58 times lower than the predicted value. Applying this scale factor to our model for the radial velocity amplitude yields $A_{\rm RV} = 1.11\pm0.09$ m s$^{-1}$. \rrev{We recognize that this is an inferred value, based on the observed residual amplitude of the time series rather than the measured amplitudes of individual modes, and that we should not expect to precisely recover the true value of $A_{\rm RV}$ given the short baseline of the NEID time series. But the discrepancy between the inferred RV oscillation amplitude and the predicted value is nevertheless significant, as we illustrate in Figure \ref{fig:rv_amp}.}

\section{Discussion}\label{sec:discussion}

\subsection{Causes of amplitude suppression}\label{sec:significance}
The oscillation amplitudes detected in both the TESS light curves and the NEID RV time series are weaker than anticipated.
This discrepancy is marginal in the photometry, at $<2\sigma$, but significant in the RV \rrev{residual amplitude} data.
To better understand the underlying cause, we first consider the precision and accuracy of the stellar parameters. We use these parameters as inputs to the scaling relations from which the predicted amplitudes are derived; if the inputs are inaccurate it is possible that the observed oscillation amplitudes are \textit{not} inconsistent with the theoretical scaling relations. But if the stellar parameters are reasonably accurate, this points to shortcomings of the scaling relations, or of our understanding of the physics of p-mode oscillations, as the reason for the observed amplitude discrepancy.

\subsubsection{Accuracy of stellar parameters}
\begin{figure*}
    \centering
    \includegraphics[width=1.0\linewidth]{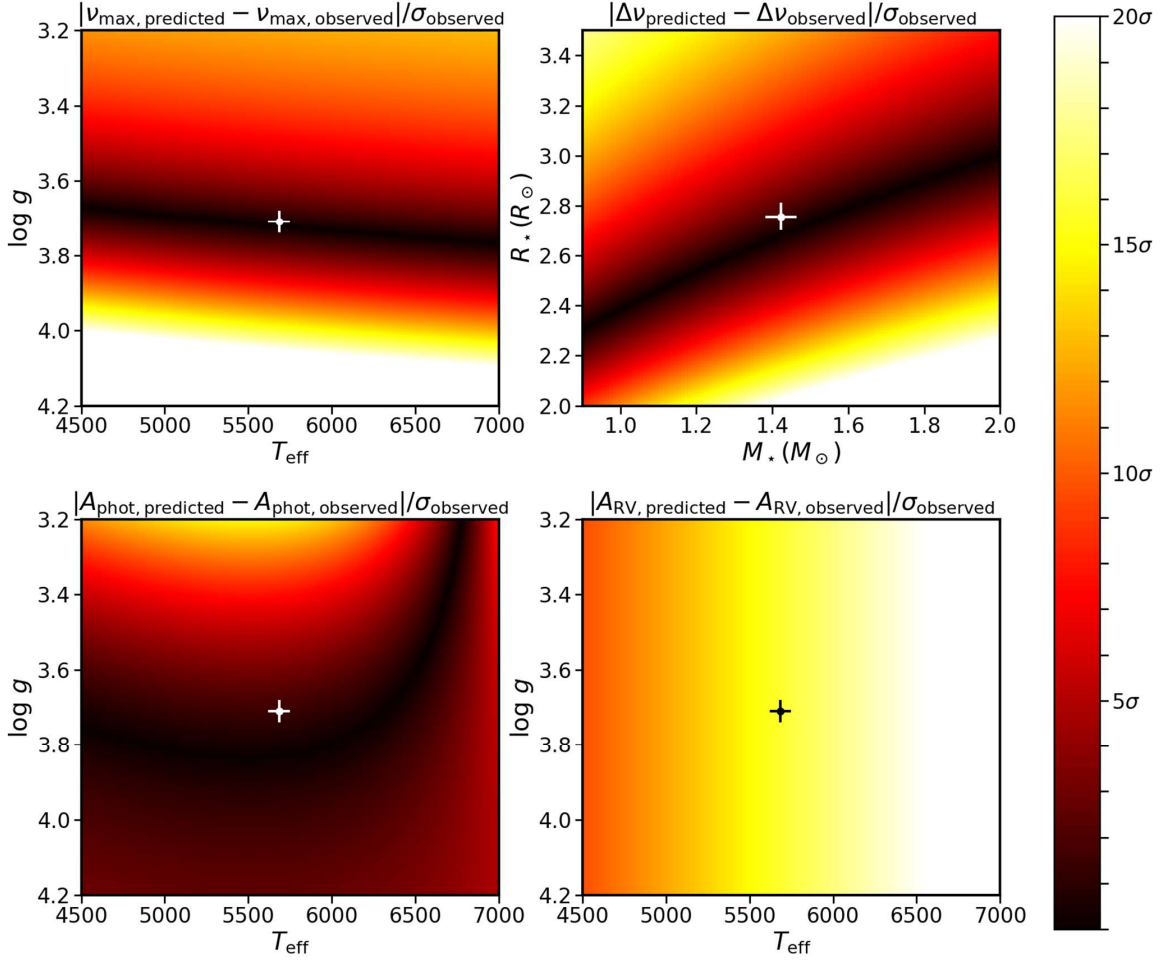}
    \caption{Relative difference between the predicted and observed values of $\nu_{\rm max}$, $\Delta \nu$, $A_{\rm phot}$, and $A_{\rm RV}$ for grids of possible stellar parameters for HD 35833. The dark swaths indicate regions of parameter space for which the predicted values would be consistent with observations. The $\teff$ and $\log g$ values ($M_\star$ and $R_\star$ values for $\Delta \nu$) given in Table \startable\ are shown as a white circle (black for $A_{\rm RV}$), with error bars for the formal uncertainties.
    We note that the observed values are consistent with predictions for $\nu_{\rm max}$ and $\Delta \nu$, and $<5\sigma$ discrepant for $A_{\rm phot}$. But there is no reasonable set of stellar parameters that is simultaneously compatible with the aforementioned values as well as with the observed value of $A_{\rm RV}$.}
    %We note that only a very narrow range of parameters is simultaneously compatible with the observed values of both $\nu_{\rm max}$ and $A_{\rm phot}$, and this region differs significantly from our best measurements of the values of $\teff$ and $\log g$.
    \label{fig:siginificance_maps}
\end{figure*}

We re-parameterize Equations \ref{eq:aphot} and \ref{eq:aphottred} in terms of $\teff$ and $\log g$ to reduce the number of free parameters in the amplitude scaling relations
\begin{equation}\label{eq:aphot_tefflogg}
    A_{\rm phot}  ={\rm A}_{\rm phot,\odot}
    \beta\left(\frac{g}{{\rm g}_\odot}\right)^{-1}\left(\frac{\teff, \star}{{\rm T}_{\rm eff, \odot}}\right)^{2}
\end{equation}
and
\begin{equation}\label{eq:aphot_tred_tefflogg}
%\begin{split}
    T_{\rm red} = 8907\left(\frac{R_\star}{{\rm R}_\odot}\right)^{-0.186}\left(\frac{\teff, \star}{{\rm T}_{\rm eff, \odot}}\right)^{-0.372} {\rm\ K}
%\end{split}
\end{equation}
where $\beta$ still depends on $T_{\rm red}$ as in Equation \ref{eq:aphotbeta}. 
The scaling relation for the RV oscillation amplitude given in Equation \ref{eq:amaxrv} can similarly be reparameterized as
\begin{equation}\label{eq:amaxrv_tefflogg}
    A_{\rm max, RV} = {\rm A}_{\rm max, RV, \odot} \left(\frac{g}{{\rm g}_\odot}\right)^{-1}\left(\frac{\teff, \star}{{\rm T}_{\rm eff, \odot}}\right)^{4}.
\end{equation}
We calculate the predicted values of $A_{\rm phot}$ and $A_{\rm RV}$ for values of $\teff$ and $\log g$ on the ranges $4500 $ K $< \teff < 7000$ K and $3.2 < \log g < 4.2$, holding the only remaining free parameter, $R_\star$, fixed to the value given in Table \startable. We show the relative difference between the predicted and observed values of $A_{\rm phot}$ in Figure \ref{fig:siginificance_maps} alongside an equivalent plot for $\nu_{\rm max}$. We also show the relative difference between the predicted and observed values of $\Delta \nu$ as a function of stellar mass and radius.
Whereas the observed values of $\nu_{\rm max}$ and $\Delta \nu$ agree well with predictions, the amplitudes do not. The RV amplitude, in particular, is not consistent to $<10\sigma$ for any reasonable set of stellar parameters. 
Furthermore, it is readily apparent that a change in $\log g$ by more than $\sim0.1$ dex would be incompatible with the observed value of $\nu_{\rm max}$. Uncertainties in the stellar parameters are therefore rather unlikely to be responsible for the apparent suppression of the oscillation amplitudes.

\subsubsection{Magnetic suppression of p-mode oscillations}\label{sec:activity}

Detailed helioseismic studies have shown that the frequencies and amplitudes of solar p-mode oscillations vary with the solar activity cycle \citep{Chaplin2000,Jimenez-Reyes2003}, with higher levels of activity correlating with weaker oscillation amplitudes. And in more recent years, studies of individual stars \citep[e.g.,][]{Garcia2010,Bonanno2019} and large stellar samples \citep{Chaplin2011,Bonanno2014,Mathur2019} have revealed this same anticorrelation in stars other than the Sun.
These findings suggest that stellar activity can lead to the suppression of p-mode oscillations, a prediction that is corroborated by solar magnetoconvection models \citep{Cattaneo2003} and empirical results from spatially resolved observations of oscillations in the vicinity of sunspots \citep{Braun1987,Braun1988}.

\begin{figure}
    \centering
    \includegraphics[width=1.05\linewidth]{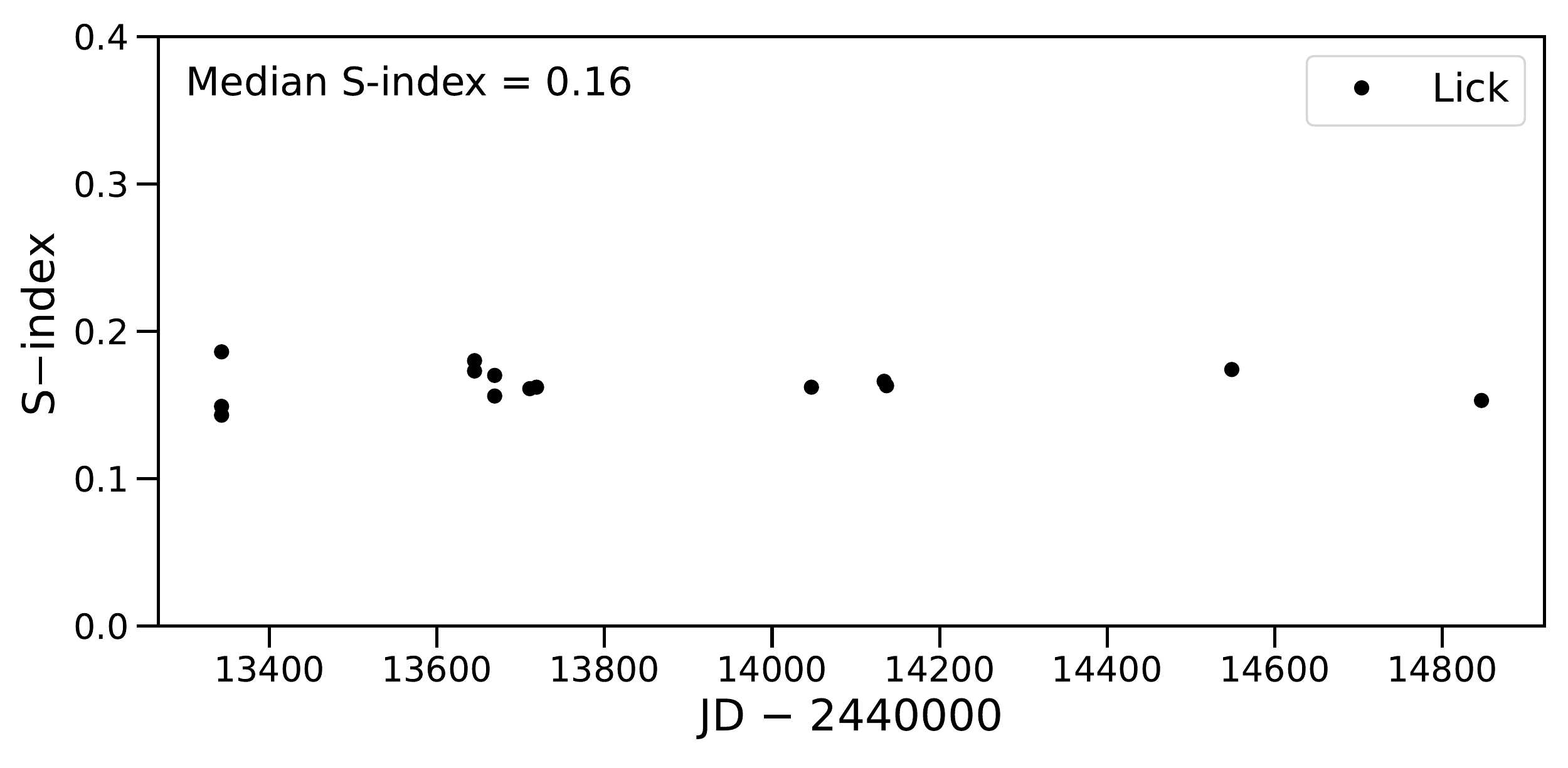}
    \caption{CaII H \& K S-index measurements for HD 35833 from Lick observations \citep{Isaacson2010}. The median value of $0.16$ shows that activity-induced suppression can not explain the observed p-mode oscillation amplitudes.}
    \label{fig:lick_activity}
\end{figure}

We use the CaII H \& K S-index as a proxy for chromospheric activity to assess whether magnetic suppression is the mechanism responsible for the weaker-than-expected RV oscillation signal observed for HD 35833. As \citet{Bonanno2014} show, high S-index values are correlated with smaller oscillation amplitudes in the \textit{Kepler} bandpass; we expect the same to be true for TESS, of course, and for NEID as well, as photometric and radial velocity oscillation signals are simply different manifestations of the same physical process. However, measurements of the S-index for HD 35833 show that this star is not magnetically active.
\citet{Isaacson2010} calculate S-index values for HD 35833 for 14 nights of Lick observatory data collected across $4.5$ years from 2004 to 2009 as part of the  California Planet Search survey; we show these measurements in Figure \ref{fig:lick_activity}. The median S-index for the Lick time series is $0.16$, and the values range from $0.14 - 0.19$. This is consistent with the values reported by \citet{Isaacson2010} for very quiet subgiants with similar $B-V$ color, indicating that HD 35833 was not exhibiting high levels of activity at the time.

\begin{figure}
    \centering
    \includegraphics[width=1.05\linewidth]{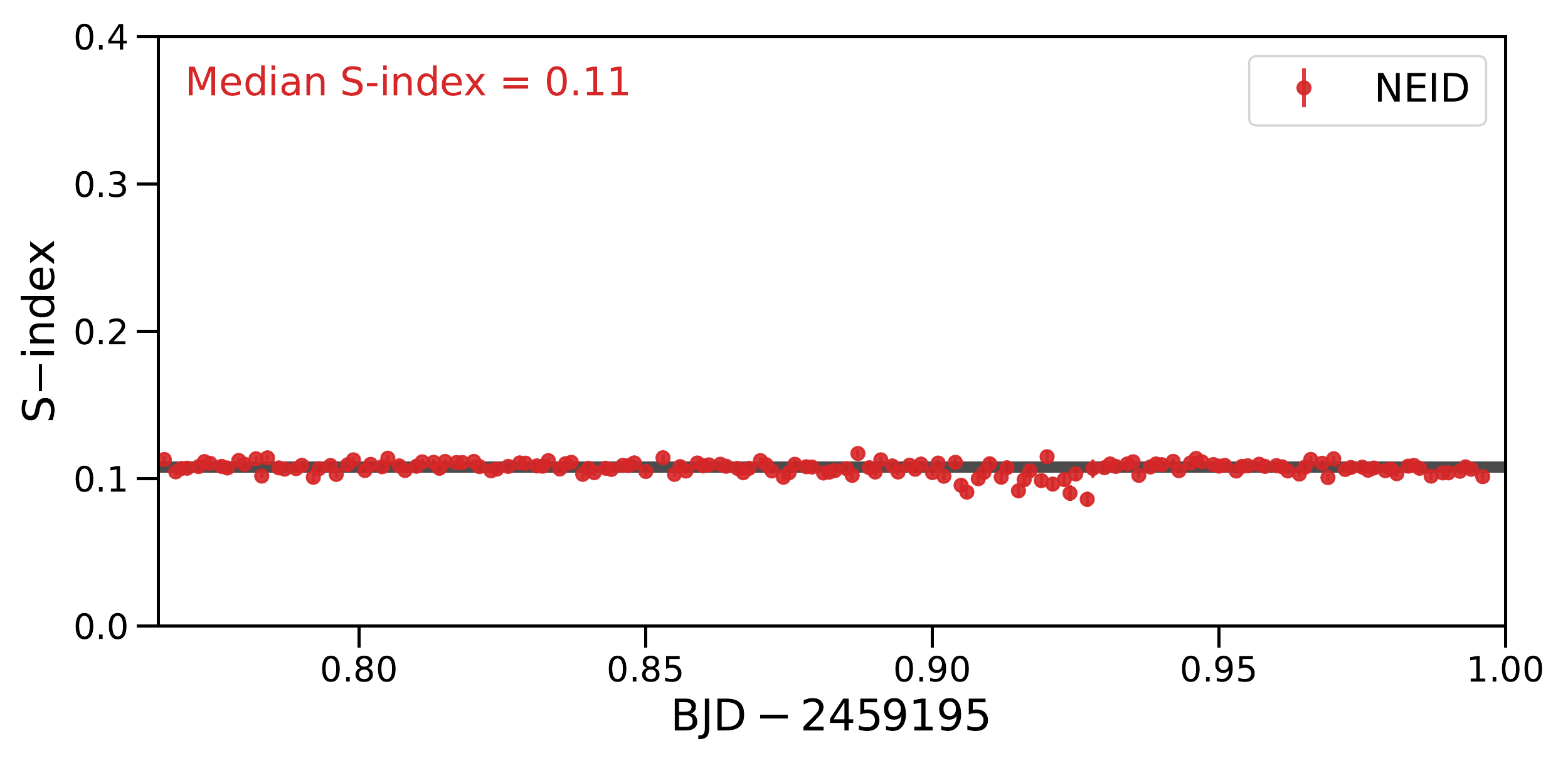}
    \caption{CaII H \& K S-index measurements for HD 35833 from the NEID time series.}
    \label{fig:activity}
\end{figure}

We also consider present day activity levels for HD 35833, as this is the more pertinent metric in the context of the NEID observations. The S-indices for the NEID time series are calculated using a modified version of the NEID DRP, in which the flux in the line cores and reference regions is determined using a weighted mean of multiple overlapping orders rather than a single order. The median S-index for the NEID observations of HD 35833 is $0.11$, with only modest variation throughout the night (Figure \ref{fig:activity}).
While these values are not calibrated to the same scale as the \citet{Isaacson2010} sample and thus cannot be directly compared, we confirm that the star remained quiet by noting the absence of emission features in the CaII H \& K line cores in the NEID spectra.
This is consistent with our expectation that activity levels for subgiants should not change significantly on decade-long timescales \citep{Wright2004}.
HD 35833 was not exhibiting high levels of activity on the night it was observed with NEID, and the oscillations were not magnetically suppressed.

\subsubsection{Other causes of amplitude suppression}\label{sec:other_supp}

\rrev{
\citet{Mathur2019} explore the possibility that metallicity may be responsible for a reduction in oscillation amplitudes in a sample of \rrevb{\textit{Kepler}} stars for which no strong magnetic activity signals are detected. A dependence on metallicity was first suggested by \citet{Houdek1999}, who point to the impact of opacity on convective velocities and find that lower metallicity stars indeed expected to have weaker oscillation amplitudes. This finding is supported by more detailed 3D hydrodynamical simulations by \citet{Samadi2010}. However, the results of the \citet{Mathur2019} analysis are inconclusive; while the stars without detected oscillations have sub-solar metallicities on average, a significant number of them have [Fe/H]$>0$. With the lack of observational evidence confirming the expected effect of metallicity, we refrain from considering it as the source of oscillation suppression in the case of HD 35833.
}

\rrev{
In using the correction factor $c$ from \citet{Kjeldsen2008} to calculate the expected oscillation amplitudes in Section \ref{sec:astero_scaling}, we implicitly assume that the relative visibilities of different modes in HD 35833 are similar to those of the Sun.
However, multiple studies have shown that the visibilities of the dipolar ($l=1$) oscillation modes may be significantly reduced or absent altogether for some evolved stars \citep{Mosser2012,Garcia2014,Stello2016}.
If these modes are not present in HD 35833, the overall oscillation amplitudes will be weaker than the predicted values given in Table \asterotable.
To determine whether this might explain the observed RV amplitude suppression, we re-calculate the predicted residual amplitude signal for the NEID time series in the extreme case that the visibility $S_1=0$. We find that while the ratio between the predicted and observed residual amplitudes is not as drastic in this case ($A_{\rm RV,predicted}/A_{\rm RV,observed}=1.97$), the discrepancy is still significant at the $>5\sigma$ level, indicating that missing dipolar modes cannot be the sole cause of the observed suppression.
}

\subsection{Implications for radial velocity exoplanet detection}

The agreement between the predicted and observed shapes of the oscillation RV RMS for HD 35833 (Figure \ref{fig:rv_amp}) empirically validates the \citet{Chaplin2019} residual amplitude model for integration times $t_i<\nu_{\rm max}^{-1}$ for the first time. This complements the \citet{Chaplin2019} analysis of $\alpha$ Cen A RVs, for which the residual amplitudes show good agreement with the model at integration times $t_i>\nu_{\rm max}^{-1}$. 
In addition, our analysis in Section \ref{sec:gp_decomp} highlights the useful application of the \citet{Luhn2022} Gaussian process kernels in fitting and removing correlated RV signals. 
These results support the use of the \citet{Chaplin2019} attenuation model and the \citet{Luhn2022} correlated noise model to inform exposure times and observing strategies for EPRV exoplanet surveys.

At the same time, we must recognize that there are gaps in our current understanding of the physical processes governing p-mode oscillations, and that we have yet to understand these signals at the 10~\cms{} level.
%Unless HD 35833 is an anomalous case
The significant discrepancy between the \rrev{residual} RV oscillation amplitudes we detect and the \rrev{residual} amplitudes predicted by scaling relations is cause for concern, as it indicates there may be additional physics not captured by these relations. Though this mechanism suppressed the oscillations in HD 35833, it may in other cases excite stronger oscillations, making it more difficult to detect exoplanets with small semi-amplitudes.
In addition, the TESS and NEID data for HD 35833 reveal a very different ratio between photometric and RV amplitudes than scaling relations predict, suggesting that we should be more cautious about using photometric data to inform estimates of the RV noise contribution from oscillations.
Inaccurate or imprecise models of stellar variability will also inhibit robust mass and orbit measurements for any low-amplitude exoplanet signals that are detected. Even with a reliable strategy in place for mitigating the noise contribution of p-mode oscillations, we must continue to undertake detailed studies of these signals if we aim to push the limits of RV exoplanet detection and characterization.

\section{Conclusion}\label{sec:conclusion}

We report separate detections of p-mode oscillations in the subgiant star HD 35833 with NEID and TESS. Independent analyses of both data sets show that the oscillation amplitudes are significantly weaker than expected from scaling relations, yet we are unable to link this finding to any known suppression mechanisms. 
The NEID data also validate the \citet{Chaplin2019} amplitude attenuation model for RV observations with exposure times $>\nu_{\rm max}^{-1}$, resolving a critical unknown in the design of current and future EPRV exoplanet surveys.

We failed to detect oscillations simultaneously in RVs and photometry, because the elevated white noise floor in the Sector 32 TESS data precluded the detection of the p-mode signal. While these data were not wholly uninformative, as upper limits on the oscillation amplitude in Sector 32 are consistent with predicted amplitudes, future analyses in this same vein would still be valuable.
If a tight relation between photometric and RV oscillation amplitudes can be identified, observers can develop more informed RV observing strategies by using long-baseline photometric data (e.g., from TESS or its successor) to more accurately predict the strength of the oscillaiton signal in RVs. We also invite readers to make use of the data sets presented herein for future studies of oscillations involving larger ensembles of stars.

\section{Acknowledgements}

NEID is funded by NASA through JPL by contract 1547612 and the NEID Data Reduction Pipeline is funded through JPL contract 1644767. We acknowledge support from the Heising-Simons Foundation via grant 2019-1177.
The Center for Exoplanets and Habitable Worlds and the Penn State Extraterrestrial Intelligence Center are supported by the Pennsylvania State University and the Eberly College of Science.
This research has made use of the SIMBAD database, operated at CDS, Strasbourg, France, and NASA's Astrophysics Data System Bibliographic Services.
Part of this work was performed for the Jet Propulsion Laboratory, California Institute of Technology, sponsored by the United States Government under the Prime Contract 80NM0018D0004 between Caltech and NASA.

\rrev{This paper contains data taken with the NEID instrument, which was funded by the NASA-NSF Exoplanet Observational Research (NN-EXPLORE) partnership and built by Pennsylvania State University. NEID is installed on the WIYN telescope, which is operated by the NSF's National Optical-Infrared Astronomy Research Laboratory, and the
NEID archive is operated by the NASA Exoplanet Science Institute at the California Institute of Technology.
NN-EXPLORE is managed by the Jet Propulsion Laboratory, California Institute of Technology under contract with the National Aeronautics and Space Administration. We thank the NEID Queue Observers and WIYN Observing Associates for their skillful execution of our observations.}

This work includes data collected by the TESS mission, which are publicly available from MAST. Funding for the TESS mission is provided by the NASA Science Mission directorate. Some of the data presented in this paper were obtained from MAST. Support for MAST for non-HST data is provided by the NASA Office of Space Science via grant NNX09AF08G and by other grants and contracts. This work has made use of data from the European Space Agency (ESA) mission {\it Gaia} (\url{https://www.cosmos.esa.int/gaia}), processed by the {\it Gaia} Data Processing and Analysis Consortium (DPAC, \url{https://www.cosmos.esa.int/web/gaia/dpac/consortium}). Funding for the DPAC has been provided by national institutions, in particular the institutions participating in the {\it Gaia} Multilateral Agreement.

Based in part on observations at Kitt Peak National Observatory, NSF’s NOIRLab (Prop. ID 2020B-0417; PI: A.\ Gupta), managed by the Association of Universities for Research in Astronomy (AURA) under a cooperative agreement with the National Science Foundation. The authors are honored to be permitted to conduct astronomical research on Iolkam Du’ag (Kitt Peak), a mountain with particular significance to the Tohono O’odham.
\rrev{We also express our deepest gratitude to Zade Arnold, Joe Davis, Michelle Edwards, John Ehret, Tina Juan, Brian Pisarek, Aaron Rowe, Fred Wortman, the Eastern Area Incident Management Team, and all of the firefighters and air support crew who fought the recent Contreras fire. Against great odds, you saved Kitt Peak National Observatory.}
%Data presented herein were obtained at the WIYN Observatory from telescope time allocated to NN-EXPLORE through the scientific partnership of the National Aeronautics and Space Administration, the National Science Foundation, and the National Optical Astronomy Observatory.

The Pennsylvania State University campuses are located on the original homelands of the Erie, Haudenosaunee (Seneca, Cayuga, Onondaga, Oneida, Mohawk, and Tuscarora), Lenape (Delaware Nation, Delaware Tribe, Stockbridge-Munsee), Shawnee (Absentee, Eastern, and Oklahoma), Susquehannock, and Wahzhazhe (Osage) Nations.  As a land grant institution, we acknowledge and honor the traditional caretakers of these lands and strive to understand and model their responsible stewardship. We also acknowledge the longer history of these lands and our place in that history.

\facilities{TESS, WIYN (NEID), \textit{Gaia}}

\software{astropy \citep{AstropyCollaboration2018}, barycorrpy \citep{Kanodia2018}, lightkurve \citep{LightkurveCollaboration2018}, matplotlib \citep{Hunter2007}, numpy \citep{Harris2020}, scipy \citep{Oliphant2007}, SERVAL \citep{Zechmeister2018}}

\bibliography{references}{}
\bibliographystyle{aasjournal}

\appendix

\section{Residual radial velocity amplitudes for sequences of exposures}

\begin{figure}[ht]
    \centering
    \includegraphics[width=0.95\linewidth]{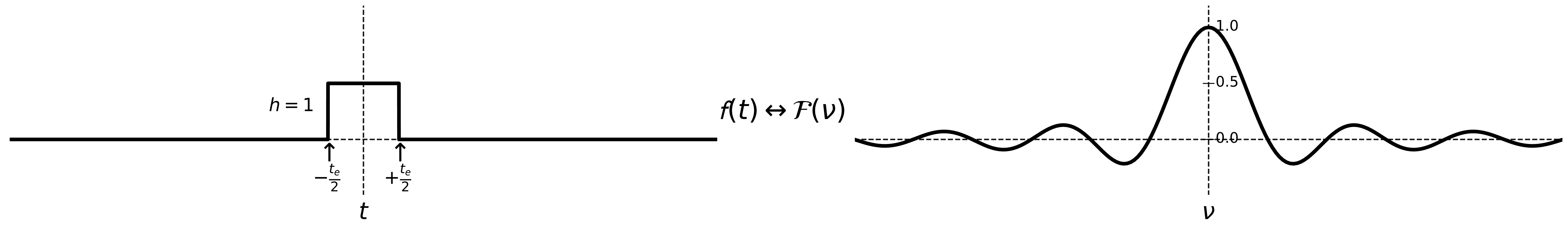}
    \caption{Transfer function for a single finite exposure.}
    \label{fig:single_sinc}
\end{figure}

\begin{figure}[ht]
    \centering
    \includegraphics[width=0.95\linewidth]{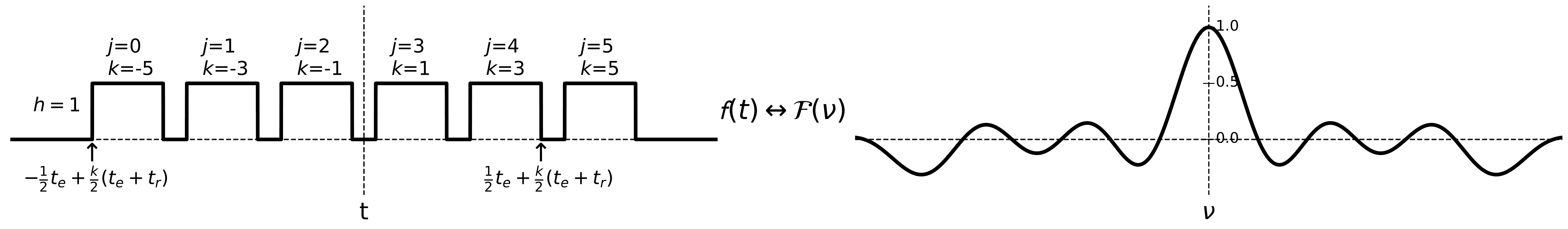}
    \caption{Transfer function for a sequence of 6 exposures separated by non-zero readout time.}
    \label{fig:multi_sinc}
\end{figure}

Here, we describe the analytic calculation of the residual RV amplitude of a stellar oscillation signal for a sequence of exposures separated in time. We begin with the simple case of a single exposure, and then generalize this to multiple exposures.
For a single, continuous observation of duration $t_e$, the attenuation of the oscillation signal, or the transfer function, is given by the Fourier transform of a rectangular pulse centered on zero with width $t_e$ and unit height (Figure \ref{fig:single_sinc}).
\begin{equation}
    \begin{split}
    \eta(\nu) & = \frac{1}{t_e}\int_{-\frac{1}{2}t_e}^{\frac{1}{2}t_e}e^{-2\pi i \nu t} dt \\
     & =\frac{1}{\pi \nu t_e}\frac{e^{\pi i \nu t_e} - e^{-\pi i \nu t_e}}{2i} \\
    & = \frac{1}{\pi \nu t_e}\sin(\pi \nu t_e) \\
    & = \sinc(\pi \nu t_e). \\
    \end{split}
\end{equation}
In this case, the transfer function is equivalent to a simple sinc function.
For a sequence of observations each with exposure time $t_e$ and separated by readout time $t_r$, the transfer function is the Fourier transform of a sequence of rectangular pulses \textit{collectively} centered on zero (Figure \ref{fig:multi_sinc}).
To calculate the Fourier transform, we parameterize the phase offsets using the variable
\begin{equation}\label{eq:kparam}
        k=2j-N+1
\end{equation}
where $N$ is the total number of observations and $j$ is the index of a given observation. The transfer function is then
\begin{equation}
    \begin{split}
    \eta(\nu) & =\frac{1}{Nt_e}\sum_{j=0}^N\int_{-\frac{1}{2}t_e+\frac{k}{2}(t_e+t_r)}^{\frac{1}{2}t_e+\frac{k}{2}(t_e+t_r)}e^{-2\pi i \nu t} dt \\
    & =\frac{1}{Nt_e}\sum_{j=0}^N\frac{1}{\pi \nu} \frac{e^{\pi i \nu t_e} - e^{-\pi i \nu t_e}}{2i}e^{k\pi i\nu t_e} e^{k\pi i\nu t_r} \\
    & =\frac{1}{N}\sinc(\pi \nu t_e)\sum_{j=0}^N e^{\pi i\nu k(t_e+t_r)}. \\
    \end{split}
\end{equation}
Substituting $j$ and $N$ back in for $k$ using Equation \ref{eq:kparam}, this yields
\begin{equation}
    \eta(\nu)=\frac{1}{N}\sinc(\pi \nu t_e)\sum_{j=0}^N e^{\pi i\nu(2j-N+1)(t_e+t_r)}.
\end{equation}

%%%%%%%%%%%
%%%%%%%%%%%%%%%%%%%%%%
%%%%%%%%%%%%%%%%%%%%%%%%%%%%%%%%%
%%%%%%%%%%%%%%%%%%%%%%%%%%%%%%%%%%%%%%%%%%%%

%%%%%%%%%%%
%%%%%%%%%%%%%%%%%%%%%%
%%%%%%%%%%%%%%%%%%%%%%%%%%%%%%%%%
%%%%%%%%%%%%%%%%%%%%%%%%%%%%%%%%%%%%%%%%%%%%

\end{document}